\pdfoutput=1
\documentclass[cernpreprint, atlasdraft=false, UKenglish, texmf, orcidlogo]{atlasdoc}
\usepackage{atlaspackage}

\usepackage{siunitx}
\usepackage{atlasbiblatex}
\usepackage{feynmp}
\usepackage{bm}
\usepackage{graphicx}
\usepackage{amsmath}
\usepackage{float}
\usepackage{atlasphysics}
\graphicspath{{logos/}{figures/}}

\usepackage{ANA-HMBS-2024-28-PAPER-defs}

\AtlasTitle{Search for dimuon resonance in the 35 to 75 GeV mass range using 140~fb$^{-1}$ of 13 TeV $pp$ collisions with the ATLAS
detector}

\AtlasAbstract{%
A model-independent search for low-mass resonances decaying into pairs of oppositely charged muons is presented. The analysis uses proton--proton collision data corresponding to an integrated luminosity of 140~fb$^{-1}$, recorded by the ATLAS detector at the Large Hadron Collider between 2015 and 2018. The search targets hypothetical dimuon resonances in the invariant mass range from 35 GeV to 75 GeV. The modelling of this mass region is particularly challenging for conventional analytic background parameterisations. To address this, a Gaussian process regression technique is used to model the background. The dimuon mass spectrum is analysed for potential signals, and no statistically significant excess is observed. Upper limits at the 95\% confidence level are set on the fiducial production cross-section of new resonances decaying promptly into muons, ranging from 20~fb to 110~fb, depending on the resonance mass. These results are further interpreted in the context of dark-photon and dark-matter-mediator models, leading to new constraints on their parameter spaces.
}

\AtlasRefCode{HMBS-2024-28}
\PreprintIdNumber{CERN-EP-2025-285}

\AtlasJournalRef{JHEP 06 (2026) 165}
\AtlasDOI{https://doi.org/10.1007/JHEP06(2026)165}
\AtlasJournal{JHEP}
\hypersetup{pdftitle={ATLAS document},pdfauthor={The ATLAS Collaboration}}

\begin{document}

\maketitle

\tableofcontents

%
%

%

% The next lines are included from the .//sections/introduction.tex input file
\section{Introduction}
Many strategies are employed to search for physics beyond the Standard Model (BSM),
motivated by the open questions that remain in particle physics.
Model-independent searches for low-mass resonances are of considerable interest due to their potential to reveal new phenomena, such as possible dark-matter interactions. Among the possible final states, the dimuon channel provides a clean and well-understood
signature, offering high sensitivity to a broad class of BSM scenarios.

BSM scenarios involving dark matter (DM) provide a compelling motivation for such searches. While there is convincing cosmological evidence for the existence of dark matter~\cite{Bertone2005,Feng:2010gw,Porter:2011nv,Akrami:2018vks}, its fundamental nature remains one of the most pressing open questions in both particle physics and cosmology. Although no interactions between DM and Standard Model (SM) particles beyond gravity have yet been observed, many theoretical models predict that DM could interact very weakly with SM particles~\cite{EXOT-2017-32}. These scenarios often introduce a hidden sector of particles, with interactions mediated by a new vector or axial-vector boson. If such interactions exist, these mediators could be produced in proton--proton ($pp$) collisions at the Large Hadron Collider (LHC)~\cite{Evans:2008zzb} and subsequently decay into SM particles.

While the search presented in this paper is designed to be model independent, two benchmark theoretical scenarios are used for interpretation, illustrating possible DM realisations leading to low-mass dimuon resonances.  A standard benchmark for LHC DM searches~\cite{ABERCROMBIE2020100371} involves extending the SM with an additional U(1) gauge symmetry, under which the DM particle carries a charge. If certain SM particles are also assumed to carry charges under this new gauge group, a new gauge boson, commonly denoted as $Z^\prime$, could mediate interactions between SM and DM particles. Within the framework of a simplified $s$-channel exchange model, these interactions can be described by the following Lagrangians, incorporating both vector and axial-vector couplings:
\begin{align}
\label{eq:Zp_interactions}
\mathcal{L}_V &\supset \frac{1}{2} m_{Z'}^2 V_{\mu} V^{\mu} - g_{q} V_{\mu} \bar{q}_i \gamma^{\mu} q_i - g_{l} V_{\mu} \bar{l}_i \gamma^{\mu} l_i - g_{\chi} V_{\mu} \bar{\chi} \gamma^{\mu} \chi\\
\label{eq:Zp_interactions2}
\mathcal{L}_A &\supset \frac{1}{2} m_{Z'}^2 V_{\mu} V^{\mu} - g_{q} V_{\mu} \bar{q}_i \gamma^{\mu} \gamma^5 q_i - g_{l} V_{\mu} \bar{l}_i \gamma^{\mu} \gamma^5 l_i - g_{\chi} V_{\mu} \bar{\chi} \gamma^{\mu} \gamma^5 \chi
\end{align}
where $m_{Z'}$ is the mass of the $Z^{\prime}$ boson ($V_\mu$). The couplings $g_q$ and $g_l$ are taken to be universal across all quark and lepton flavours, respectively, and $g_\chi$ represents the coupling between the mediator and the DM particle ($\chi$).

An alternative, theoretically well-motivated framework introduces kinetic mixing between a dark Abelian gauge symmetry, $\text{U}_\text{D}$(1), and the SM hypercharge, $\text{U}_Y$(1)~\cite{Curtin:2014cca}. The relevant gauge terms in the Lagrangian are:
\begin{align*}
\mathcal{L} \supset -\frac{1}{4} B_{\mu\nu} B^{\mu\nu}
- \frac{1}{4} Z_{\text{D}\mu\nu} Z_\text{D}^{\mu\nu}
+ \frac{1}{2} \frac{\epsilon}{\cos\theta} Z_{\text{D}\mu\nu} B^{\mu\nu}
+ \frac{1}{2} m_{Z_\text{D},0}^2 Z_\text{D}^\mu Z_{\text{D}\mu}.
\end{align*}
where the $\text{U}_Y$(1) and $\text{U}_\text{D}$(1) field strengths are defined as $B_{\mu\nu} = \partial_\mu B_\nu-\partial_\nu B_\mu$ and $Z_{\text{D}\mu\nu} = \partial_\mu Z_{\text{D}\nu}-\partial_\nu Z_{\text{D}\mu}$, respectively, and $\theta$ is the Weinberg mixing angle. In this model, the resonance, referred to as a dark photon, $Z_\text{D}$, has a mass $m_{Z_\text{D}}$, denoted as $m_{Z_\text{D},0}$ before mixing with the SM. While the dark photon has no direct couplings to SM particles, the kinetic mixing parameter $\epsilon$ induces an effective coupling between the two sectors.

\begin{figure}[ht]
\centering
\subfloat[]{
\centering
\includegraphics[width=0.4\linewidth]{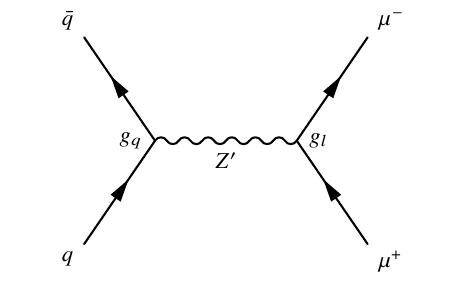}
}
\hspace{0.1\textwidth}
\subfloat[]{
\centering
\includegraphics[width=0.4\linewidth]{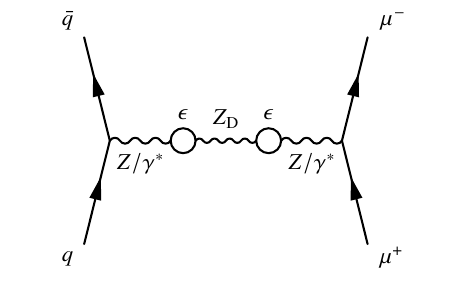}
}

\caption{Feynman diagrams for Drell--Yan production of (a) a generic resonance \(Z'\) and (b) a kinetically mixed dark photon \(Z_\text{D}\).}
\label{fig:FeynZD}
\end{figure}

This analysis searches for new resonances decaying into pairs of muons, using the dimuon invariant mass as the primary discriminating variable. Benchmark signal models include a $Z^\prime$ and a $Z_\text{D}$, which could be directly produced via the Drell–Yan (DY) mechanism in $pp$ collisions and subsequently decay promptly into a pair of oppositely charged muons, as illustrated in Figure~\ref{fig:FeynZD}. These benchmark signals serve as representative scenarios for evaluating the sensitivity of the search.

Searches for low-mass dimuon resonances have been performed by the
CMS~\cite{CMS-EXO-19-018,CMS-EXO-21-005} and LHCb~\cite{LHCb:2019vmc} Collaborations,
covering mass ranges of 1.1--7.9~\GeV, 11.5--75~\GeV and 110--200~\GeV
for CMS, and 0.214--70~\GeV\ for LHCb. Building on similar strategies, this analysis employs a data-driven approach to estimate the dominant, smoothly varying background from $Z/\gamma^*$ induced dimuon production and to search for new resonances appearing as localised deviations on top of this background spectrum using ATLAS data~\cite{PERF-2007-01}. The mass region from 35 to 75~\GeV is investigated. This range is selected based on signal-extraction capability and is explored for the first time in a resonance search with ATLAS data.

A key innovation of this study is the use of Gaussian process regression (GPR)
\cite{books/lib/RasmussenW06,frate2017modeling,Gandrakota:2022wyl},
a flexible non-parametric method that is novel in the context of modelling the background shape in resonance searches. This replaces the traditional analytic functions used, for example, in
Refs.~\cite{HIGG-2019-23,EXOT-2019-01,EXOT-2018-08,CMS-EXO-19-018,CMS-EXO-21-005,LHCb:2019vmc}, to describe the background in simultaneous signal-plus-background fits, and provides a robust description of the full mass range accessible with standard, unprescaled muon triggers.

The signal is extracted through a binned simultaneous profile likelihood fit to the signal and background components observed in data.  In the absence of a significant excess, upper limits are set on the fiducial production cross-section times branching ratio to the dimuon final state of a new resonance, as a function of its mass. These model-independent fiducial cross-sections are then parameterised into the coupling parameters of the DM mediator and dark photon models, described above, from which upper limits on these theoretical parameters are derived.

The structure of this paper is as follows. Section ~\ref{sec:detector} provides a brief description of the ATLAS detector. Section~\ref{sec:samples} presents the data and Monte Carlo (MC) simulation samples used in the analysis. The event reconstruction and selection criteria are detailed in Section~\ref{sec:eventselection}. Section~\ref{sec:signalmodelling} discusses the modelling of the signal and the associated systematic uncertainties. The background estimation and signal extraction based on GPR are described in Section~\ref{sec:backgroundmodelling}. Section~\ref{sec:ts} outlines the statistical interpretation and limit-setting procedure. The results and their interpretations in the context of benchmark models are presented in Section~\ref{sec:results}, and the conclusions are summarised in Section~\ref{sec:conclusions}.

% End of text imported from the .//sections/introduction.tex input file

%

%

% The next lines are included from the .//sections/atlas.tex input file
\section{ATLAS detector}
\label{sec:detector}

The ATLAS detector~\cite{PERF-2007-01} at the LHC covers nearly the entire solid angle around the collision point.\footnote{\AtlasCoordFootnote}
It consists of an inner tracking detector surrounded by a thin superconducting solenoid, electromagnetic and hadronic calorimeters,
and a muon spectrometer incorporating three large superconducting air-core toroidal magnets.

The inner-detector system (ID) is immersed in a \qty{2}{\tesla} axial magnetic field
and provides charged-particle tracking in the range of \(|\eta| < 2.5\).
The high-granularity silicon pixel detector covers the vertex region and typically provides four measurements per track~\cite{ATLAS-TDR-19,PIX-2018-001}.
It is followed by the semiconductor tracker (SCT), which usually contributes eight measurements per track.
These silicon detectors are complemented by the transition radiation tracker (TRT), which extends track reconstruction capabilities to a radial distance corresponding to \(|\eta| = 2.0\).
The TRT also provides electron identification information
based on the fraction of hits (typically 30 in total) above a higher energy-deposit threshold corresponding to transition radiation.

The calorimeter system covers \(|\eta| < 4.9\).
Within \(|\eta|< 3.2\), electromagnetic calorimetry is provided by barrel and
endcap high-granularity lead/liquid-argon (LAr) calorimeters,
with an additional thin LAr presampler covering \(|\eta| < 1.8\)
to correct for energy loss in material upstream of the calorimeters.
Hadronic calorimetry is provided by the steel/scintillator-tile calorimeter within \(|\eta| < 1.7\), and two copper/LAr hadronic endcap calorimeters in $1.5 < |\eta| < 3.2$.
The solid angle coverage is completed with forward copper/LAr and tungsten/LAr calorimeter modules
optimised for electromagnetic and hadronic energy measurements, respectively.

The muon spectrometer (MS) comprises separate trigger and
high-precision tracking chambers measuring the deflection of muons in a magnetic field generated by the superconducting air-core toroidal magnets.
The field integral of the toroids ranges between \num{2.0} and \qty{6.0}{\tesla\metre}
across most of the detector.
Three layers of precision chambers, each consisting of layers of monitored drift tubes, cover the region \(|\eta| < 2.7\),
complemented by cathode-strip chambers in the forward region, where the background is highest.
The muon trigger system covers the range \(|\eta| < 2.4\) with resistive-plate chambers in the barrel, and thin-gap chambers in the endcap regions.

Events were selected by the first-level trigger system implemented in custom hardware,
followed by selections made by algorithms implemented in software in the high-level trigger~\cite{TRIG-2016-01}.
The first-level trigger accepted events from the \qty{40}{\MHz} bunch crossings at a rate close to \qty{100}{\kHz},
which the high-level trigger further reduced in order to record complete events to disk at about \qty{1.25}{\kHz}.

A software suite~\cite{SOFT-2022-02} is used in data simulation, in the reconstruction and
analysis of real and simulated data, in detector operations, and in the trigger and data acquisition systems
of the experiment.

% End of text imported from the .//sections/atlas.tex input file

%
%

%

% The next lines are included from the .//sections/samples.tex input file
\section{Data and Monte Carlo samples}
\label{sec:samples}
\subsection{Data sample}

The data were recorded by the ATLAS detector during the period 2015–2018. After selecting data-taking periods corresponding to good detector conditions~\cite{DAPR-2018-01} and applying standard data-quality requirements, the total integrated luminosity amounts to 140 fb$^{-1}$~\cite{DAPR-2021-01, Avoni_2018}. Events were selected using unprescaled single-muon and dimuon triggers~\cite{TRIG-2018-01}.

For single-muon triggers, two types of trigger were employed, one with an isolation requirement and one without.  The single-muon trigger with the isolation condition selected muons with transverse momentum  $p_\mathrm{T} > 26$~\GeV that satisfied medium identification and isolation criteria~\cite{MUON-2018-03}. To enhance acceptance for high-momentum muons, an additional non-isolated single-muon trigger requiring $p_\mathrm{T} > 50$~\GeV was also used, avoiding the isolation requirement
whose efficiency decreases at high $p_\mathrm{T}$.

Two dimuon triggers were employed, one with symmetric and the other with asymmetric $p_\mathrm{T}$ requirements on the two muons. The symmetric trigger required two muons with $p_\mathrm{T} > 14$~\GeV, while the asymmetric configuration required one muon with $p_\mathrm{T} > 22$~\GeV and the other with $p_\mathrm{T} > 8$~\GeV. Neither of the dimuon triggers applied isolation requirements to the muons.

\subsection{Simulation of signal processes}
\label{sec:mc_samples}
Dedicated MC samples were produced for the benchmark signal models involving the
\(Z'\) and dark photon (\(Z_\text{D}\)) resonances. These samples are used to model the signal acceptance, efficiency, and invariant-mass distributions. The simplified DM model was used for the MC production of the $Z^\prime$ signal samples, in which the $Z^\prime$ boson couples to SM and DM particles through either vector or axial-vector interactions, as defined in Eqs.~(\ref{eq:Zp_interactions}) and (\ref{eq:Zp_interactions2}). The two interaction types are denoted as $Z_V^\prime$ and $Z_A^\prime$, respectively. To study one of the most commonly considered scenarios~\cite{EXOT-2017-32}, the mass of the DM particle is set to 10~\TeV, thereby kinematically forbidding the decay of the $Z^\prime$ boson into DM particles.

The simulations are based on a \textit{Universal FeynRules Output} (UFO)~\cite{DEGRANDE20121201,Alloul:2013bka}
implementation of the DMsimp model~\cite{Backovic:2015soa}, generated using
FeynRules~\cite{Alloul:2013bka}.
The original model was modified to include couplings to leptons~\cite{Albert:2017onk}.
Signal events were generated at leading order (LO) using \MADGRAPH[5]~\cite{Curtin:2014cca} with the A14 set of tuned parameters (tune)~\cite{ATL-PHYS-PUB-2014-021} and the \NNPDF[2.3lo]~\cite{Watt:2012tq} parton distribution function (PDF) set, followed by parton showering with \PYTHIA[8]~\cite{Sjostrand:2014zea}

The differences in acceptance and mass shape modelling between the $Z^{\prime}_{A}$ and $Z^\prime_{V}$ models are found to be negligible. Therefore, signal sample production focuses on the axial-vector interaction case, while results for the vector case can be obtained by reinterpretation, using the relationship between couplings and cross-sections.

To explore different regions of parameter space, MC samples were generated over a range of $Z^\prime$ masses and intrinsic widths. For mass modelling, benchmarks were set between 35~\GeV to 75~\GeV, in 5~\GeV intervals, with both $g_l$ and $g_q$ set to 0.1. To study intrinsic width effects, two representative mass points (45~\GeV and 65~\GeV) were chosen, and the coupling strength was varied from 0.1 to 0.3, maintaining $g_l\hspace{0.75mm} =\hspace{0.75mm} g_q$. This corresponds to intrinsic-width ratios of $\Gamma/m_{Z^\prime}\approx 0.4\%-4\% $.

The spacing between simulated mass points is validated as sufficient for interpolating the signal modelling, enabling a continuous parameterisation across the mass and width ranges of interest. The upper range of the coupling scan is chosen such that the intrinsic width becomes comparable to the detector resolution, allowing an evaluation of width effects on the analysis sensitivity. The intrinsic width $\Gamma$ used is calculated according to theoretical predictions for $Z^{\prime}_{A}$~\cite{Albert:2017onk}, which account for contributions from decays into leptons, quarks, and neutrinos, given by:
\begin{align*}
\Gamma^{q\bar{q}} & =  \frac{g_q^2 \, m_{Z^\prime}}{4\pi}
\left ( 1-4 \hspace{0.25mm} z_q \right ) ^{3/2} \, , \\
\Gamma^{\ell\bar{\ell}} & =  \frac{g_\ell^2 \, m_{Z^\prime}}{12\pi}
\left ( 1-4 \hspace{0.25mm} z_\ell \right ) ^{3/2} \, , \\
\Gamma^{\nu\bar{\nu}} & = \frac{g_\ell^2}{24\pi} m_{Z^\prime} \, ,
\end{align*} where $z_i = m_i^2/m_{Z^\prime}^2$ with $i=q,\ell$.

LO $Z_\text{D}$ samples were generated between 35~\GeV to 75~\GeV, in 5~\GeV intervals, using the hidden Abelian Higgs model (\textsc{HAHM v3})~\cite{Curtin:2014cca} implemented in \MADGRAPH[5], in order to quantify the results in terms of the dark-photon interpretation. The input parameters for \MADGRAPH[5] were derived from the chosen $Z_\text{D}$ mass ($m_{Z_\text{D}}$), together with the corresponding width ($\Gamma_{Z_\text{D}}$) and branching ratio ($\mathcal{B}_{\ell\ell}$), calculated including three-loop quantum chromodynamics (QCD) corrections as reported in Ref.~\cite{Curtin:2014cca}. In the (\textsc{HAHM v3}) configuration used in this analysis, no hidden-sector states lighter than the $Z_\text{D}$ are assumed, and therefore the $Z_\text{D}$ decays exclusively into Standard Model fermions. The \NNPDF[2.3lo]~\cite{Watt:2012tq} PDF set was used for the matrix-element calculation. Parton showering and hadronisation were performed by \PYTHIA~8~\cite{Sjostrand:2007gs}, using the A14 tune and the \NNPDF[2.3lo]~\cite{Ball:2012cx} PDF set. Signal samples for both models were simulated using the ATLAS fast simulation framework (AtlFast-II)~\cite{SOFT-2010-01}.

For both the $Z'$ and $Z_\text{D}$ signals, cross-sections were evaluated at LO and corrected to next-to-next-to-leading order (NNLO) using a common k-factor~\cite{AH:2020cok}.

The effect of multiple interactions in the same and neighbouring bunch crossings (\pileup) for all simulated signal and background (described below) samples was modelled by overlaying the simulated hard-scattering events with inelastic $pp$ events generated with \PYTHIA[8.186]~\cite{Sjostrand:2007gs}, the \NNPDF[2.3lo] PDF set and the A3 tune~\cite{ATL-PHYS-PUB-2016-017}.
Simulated events were reweighted to match the pile-up conditions in data.
All samples were reconstructed and processed with the same trigger and reconstruction algorithms as used in data.

\subsection{Simulation of background processes}
\label{subsec:bkgsamples}
To validate the statistical analysis for the data-driven background estimation, simulated SM background processes are used. The dominant contribution arises from DY $Z/\gamma^* \rightarrow \mu^+\mu^-$ production, with smaller contributions from $Z/\gamma^* \rightarrow \tau^+\tau^-$, top-quark pair ($t\bar{t}$) production, and diboson processes.
\NewDocumentCommand{\Comix}{ o }{\AtlasMC[#1]{Comix}}
Background events from the DY $Z/\gamma^* \to\mu^+\mu^-$ and $Z/\gamma^* \rightarrow \tau^+\tau^-$ process
were simulated with
\SHERPA[2.2.1]~\cite{Bothmann:2019yzt} employing next-to-leading-order (NLO) accurate matrix elements for up to two partons, and LO-accurate matrix elements
for up to four partons. These were calculated with the \Comix~\cite{Gleisberg:2008fv}
and \OPENLOOPS~\cite{Cascioli:2011va,Denner:2016kdg} libraries and used the
\NNPDF[3.0NNLO] PDF set~\cite{Ball:2014uwa}. Parton showering was performed using the
\SHERPA parton shower~\cite{Schumann:2007mg} with MEPS@NLO
matching~\cite{Hoeche:2011fd,Hoeche:2012yf,Catani:2001cc,Hoeche:2009rj}.
Diboson processes ($WW$, $WZ$, and $ZZ$) were simulated with a similar set-up in \SHERPA[2.2.1]. The
$t\bar{t}$ samples were generated at NLO using
\POWHEGBOX~\cite{Frixione:2007nw,Re:2010bp} with the \NNPDF[3.0NLO] PDF
set interfaced to \PYTHIA[8] for parton showering and hadronisation with
the A14 tune. All background MC samples were processed by the full ATLAS detector simulation based on \GEANT~\cite{Agostinelli:2002hh}.

Typically, the equivalent integrated luminosity of the fully simulated background samples is 5–20 times higher than that of the data. To validate the signal extraction procedure with a smooth background template, an additional large DY sample was produced using a
generator-smearing technique, following the method used in the ATLAS
\(H\to\mu^+\mu^-\) analysis~\cite{HIGG-2019-14}. In this technique, generator-level muons are smeared using parameterised detector resolutions that
include non-Gaussian tails to emulate reconstruction effects.
The generator-smearing simulation was performed with \PYTHIA[8.244]. Only $Z\to\mu^+\mu^-$ and $Z\to\tau^+\tau^-$ decays were enabled. To speed up event generation without affecting the final-state muon kinematics, multi-parton interactions and hadronisation were disabled. The A14 tune, together with the \NNPDF[2.3LO] PDF set, was used as the baseline. Final-state radiation photons were simulated using the default \PYTHIA[8] QED parton shower. The equivalent integrated luminosity of the generator-smearing sample corresponds to approximately 30~ab$^{-1}$. Residual mismodelling in the generator-smearing sample was corrected by reweighting its mass spectrum to match that of the fully simulated DY sample, using a high-order polynomial to avoid propagating statistical uncertainties from the fully simulated sample.

% End of text imported from the .//sections/samples.tex input file

%

%

% The next lines are included from the .//sections/eventselection.tex input file
\section{Object and event selection}
\label{sec:eventselection}
Proton–proton interaction vertices are reconstructed from events containing at least two tracks associated with the ID, each with transverse momentum $p_\mathrm{T} > 0.5$~\GeV. Track reconstruction in the ID is performed using a combinatorial Kalman filter algorithm, seeded by hit triplets in the pixel and silicon microstrip detectors, with additional recovery of tracks formed from segments in the TRT\@. The primary hard-scatter vertex is defined as the vertex with the largest
$\sum p_\mathrm{T}^2$ of its associated tracks~\cite{PERF-2015-08}.

Most muon candidates are reconstructed using a combined track from the ID and the MS\@. To increase the muon reconstruction efficiency within the region $|\eta| < $1, where the MS coverage is limited, additional muon candidates are reconstructed by matching an ID track to either calorimeter hits consistent with a minimum-ionising particle or to a single MS track segment.

Corrections are applied to the reconstructed muon momentum in simulation to ensure precise agreement with data. These corrections to the simulated momentum scale and resolution are parameterised as an expansion in powers of the muon $p_\mathrm{T}$, with each coefficient determined as a function of $\eta$ and $\phi$~\cite{MUON-2022-01}.

Muon candidates are required to satisfy medium identification criteria~\cite{MUON-2018-03}. To reduce the contribution from fake muons originating from light-jet misidentification, a particle-flow-based isolation criterion~\cite{PERF-2015-09} is applied. In addition, to suppress muons from non-prompt heavy-flavour hadron decays, muon candidates must satisfy
$|d_0 / \sigma_{d_0}| < 3$
and
$|z_0 \sin\theta| < 0.5$~mm,
where $d_0$ is the transverse impact parameter measured relative to the measured beamline position, $\sigma_{d_0}$ is its uncertainty, and $z_0$ is the longitudinal distance between the point of closest approach in the transverse plane and the primary vertex. Muon candidates are required to have $p_\mathrm{T} > 3$~\GeV and $|\eta| < 2.4$.

Events are selected if they contain at least two muon candidates. The two muons with the highest $p_\mathrm{T}$ are chosen to form the dimuon pair and are required to have opposite electric charges.
To ensure that the trigger operated in its efficiency plateau region, dedicated selection criteria on $p_\mathrm{T}$ are applied to the muons matched to the corresponding trigger objects. These requirements are summarised in Table~\ref{tab:trigger-matching}.

The overall trigger matching~\cite{TRIG-2018-01} is performed by applying a logical OR of the criteria listed for the various trigger types in Table~\ref{tab:trigger-matching}. An event is considered matched if it satisfies the requirements of at least one trigger type. The relative contributions of different triggers depend on the signal mass: at low masses, approximately 65\% of events are selected exclusively by the asymmetric dimuon trigger, whereas at higher masses, about 61\% of events are selected by all three trigger types. The signal region is defined by the invariant mass of the dimuon pair being within the range 30--80~\GeV.

\begin{table}[htbp]
\centering
\caption{Summary of matching criteria ensuring the events are in the trigger efficiency plateau.}
\label{tab:trigger-matching}
\begin{tabular}{c|l|c}
\hline
Trigger type & Matching criteria & \( p_{\mathrm{T}} \) [\GeV] \\
\hline
Single muon & At least one muon matched to a single-muon trigger & \( >27 \) \\
\hline
Asymmetric dimuon & Leading muon matched to the high-\( p_{\mathrm{T}} \) trigger object & \( >23 \)  \\
& Subleading muon matched to the trigger low-\( p_{\mathrm{T}} \) object & \( >10 \) \\
\hline
Symmetric dimuon & Both muons matched to the two trigger objects & \( >15 \)  \\
\hline
\end{tabular}
\end{table}

% End of text imported from the .//sections/eventselection.tex input file

%

%
%
%

%

% The next lines are included from the .//sections/signalmodelling.tex input file
\section{Signal modelling}
\label{sec:signalmodelling}

Building on the event selection outlined in the previous section, the expected signal contribution is modelled to assess the sensitivity of the analysis to possible new resonances. The modelling of the signal shape, and the evaluation of the acceptance and efficiency within the fiducial region are presented in this section.

\subsection{Acceptance and efficiency}
Following the detector-level selection, the signal production cross-section is evaluated within a fiducial volume defined by applying the muon selection criteria at particle level, as summarised in Table~\ref{tab:fiducial-regions}. These criteria are defined after final-state radiation and are designed to reflect the acceptance of the corresponding trigger matching applied at the reconstruction level. Three overlapping categories, single-muon, asymmetric-dimuon and symmetric-dimuon are defined to emulate the respective trigger requirements. An event is considered to be within the fiducial region if it satisfies the conditions of at least one of these three categories.

\begin{table}[htbp]
\centering
\caption{Definition of fiducial regions.  Each event is required to contain two muons (labelled as $\mu_1$ and $\mu_2$) that satisfy all listed selection criteria.}
\label{tab:fiducial-regions}
\begin{tabular}{c|c|c|c|c}
\hline
Fiducial categories & \( p_{\mathrm{T}}^{\mu_1} \)[\GeV] & \( p_{\mathrm{T}}^{\mu_2} [\GeV]\) & \( |\eta^{\mu_1}| \) & \( |\eta^{\mu_2}| \) \\

\hline
Single muon &  > 27 &  > 3  & \( < 2.4 \) & \(  < 2.4 \) \\
\hline
Asymmetric dimuon & \( > 23 \) & \(  > 10 \) & \( < 2.4 \) & \( < 2.4 \) \\
\hline
Symmetric dimuon & \( > 15 \) & \( > 15 \) & \( < 2.4 \) & \( < 2.4 \) \\
\hline
\end{tabular}
\end{table}

The fiducial regions, defined in Table~\ref{tab:fiducial-regions}, provide the framework for evaluating signal acceptance and selection efficiency. The acceptance ($\mathcal{A}$) is defined as the fraction of all generated signal events that satisfy the fiducial selections at particle level. The efficiency ($\varepsilon$) is defined as the probability that a fiducial event also passes the full reconstruction-level selection, including all object- and event-level requirements. These quantities are used to extrapolate the measured signal yield to the fiducial phase space and to interpret the results in terms of theoretical parameters.

\begin{figure}[!h]
\centering
\subfloat[]{\includegraphics[width=0.45\textwidth]{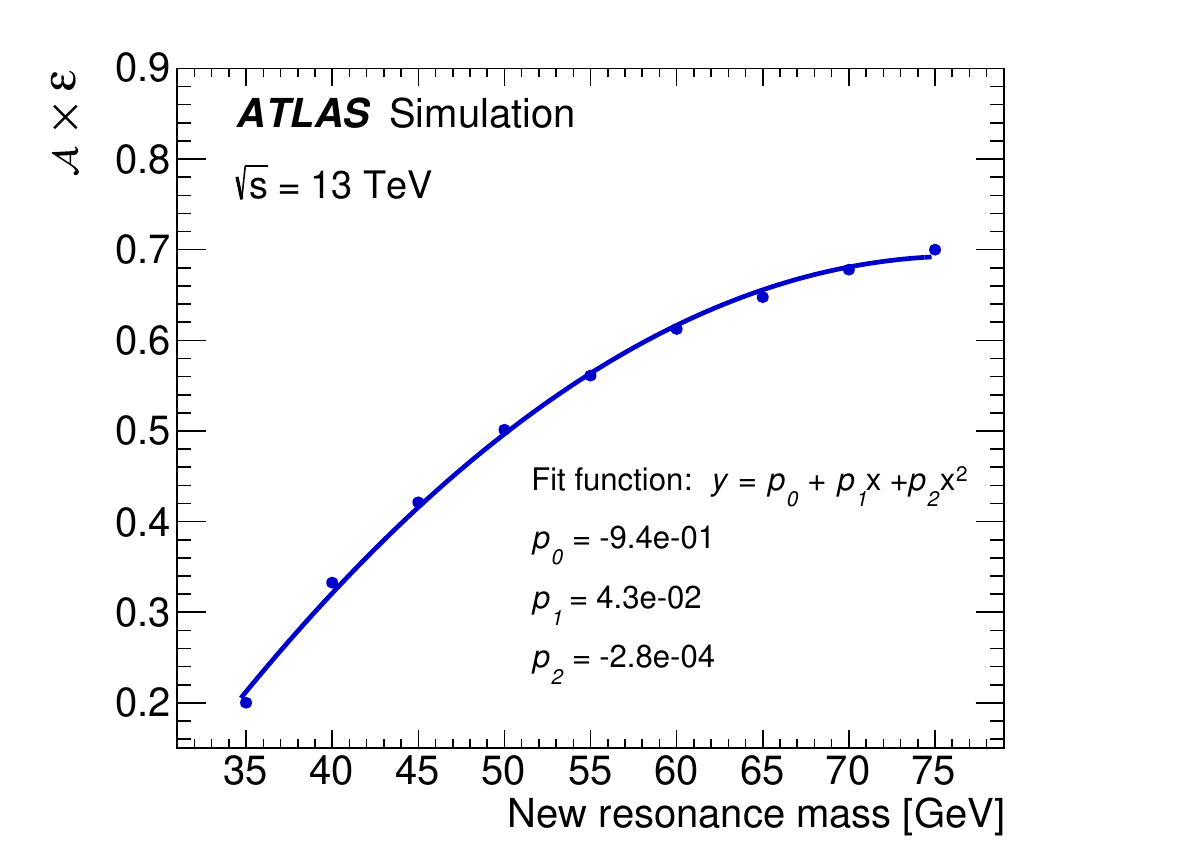}}
\subfloat[]{\includegraphics[width=0.45\textwidth]{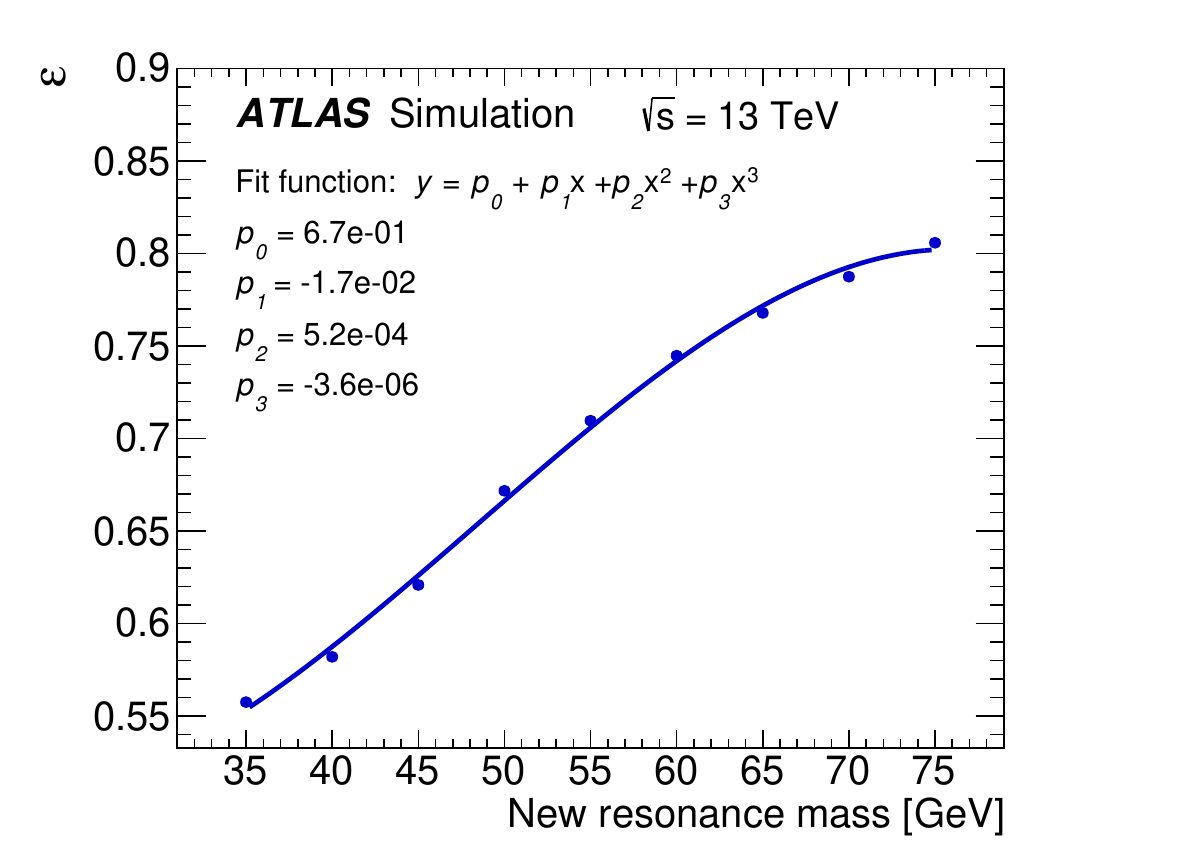}}
\caption{Parameterisation of the acceptance and efficiency for the new resonance.
(a) The product of acceptance and efficiency ($\mathcal{A} \times \varepsilon$) as a function of the resonance mass, fitted with a second-order polynomial.
(b) The detector correction factor (efficiency, $\varepsilon$) as a function of the resonance mass, fitted with a third-order polynomial.}
\label{fig:acceptance_efficiency_model}
\end{figure}

The overall acceptance and efficiency of the signal depend strongly on the mass of the new resonance. Figure~\ref{fig:acceptance_efficiency_model} presents both the acceptance times efficiency ($\mathcal{A} \times \varepsilon$) and the standalone efficiency as a function of the new resonance mass. To model their mass dependence, a second-order polynomial is fitted to $\mathcal{A} \times \varepsilon$, while a third-order polynomial is used to describe $\varepsilon$. As the acceptance and efficiency are found to be similar for the $Z^\prime$ and $Z_\text{D}$ signals, a common parameterisation is adopted for both cases.

\subsection{Parameterisation of the dimuon invariant mass}
\label{subsec:dimuonshapemodelling}
The invariant mass ($m_{\mu\mu}$) distribution of the signal resonance is modelled using a Breit--Wigner (BW) function convolved with a double-sided Crystal Ball (DSCB) function~\cite{Oreglia:1980cs, HIGG-2014-04, HIGG-2016-08}. The generator-level lineshape of the signal is described by the BW function, as:
\begin{equation*}
f(x) \propto \frac{1}{(x-m_X)^2+\left(\frac{1}{2} \Gamma \right)^2}
\end{equation*}

where $m_X$ is the new resonance mass and $\Gamma$ is the intrinsic width, which is described in Section~\ref{sec:samples}. The DSCB function is used to model the detector resolution, featuring a Gaussian distribution core with power-law tails on both sides.

In the dark photon model, the intrinsic width is expected to be sufficiently small to be negligible~\cite{Curtin:2014cca}. Therefore, the narrow width approximation (NWA) is used to test this model.

\begin{figure}[htb]
\centering
\subfloat[]{\includegraphics[width=0.49\linewidth]{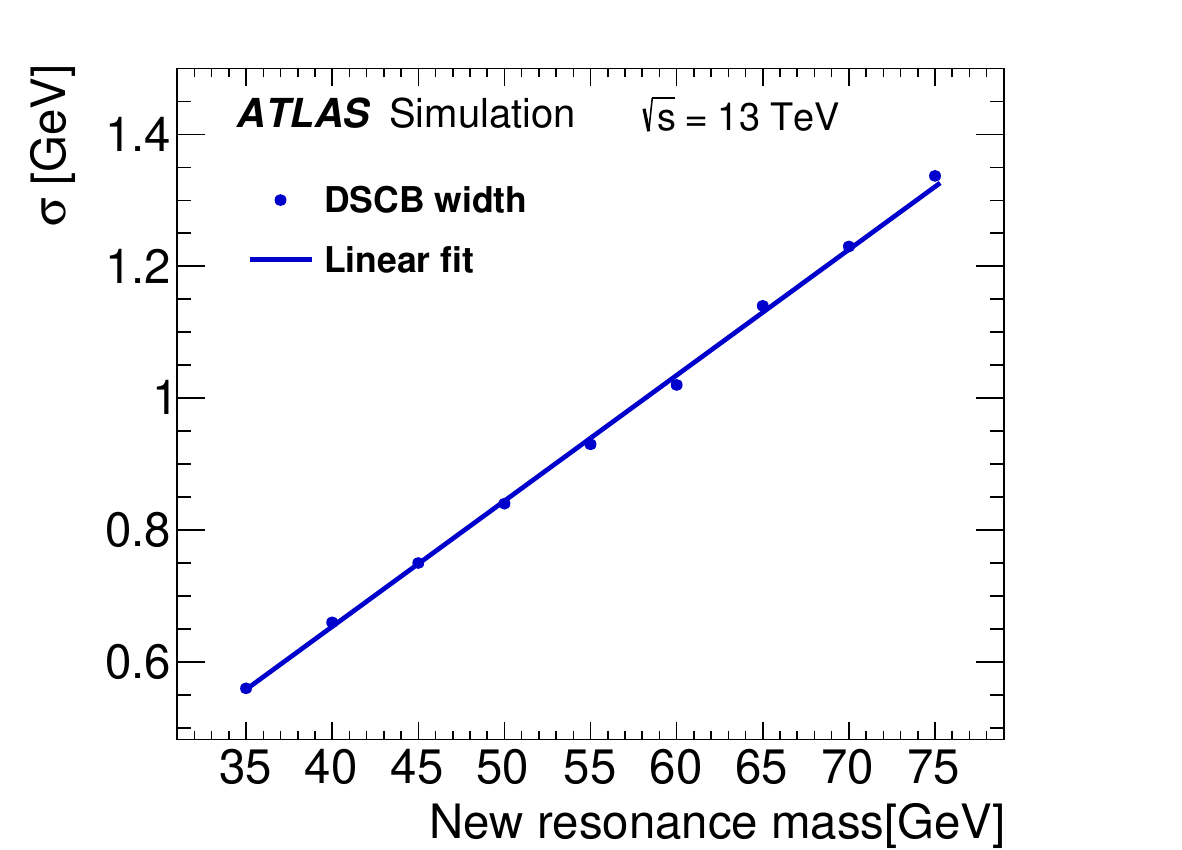} \label{fig:resolution}}
\subfloat[]{\includegraphics[width=0.49\linewidth]{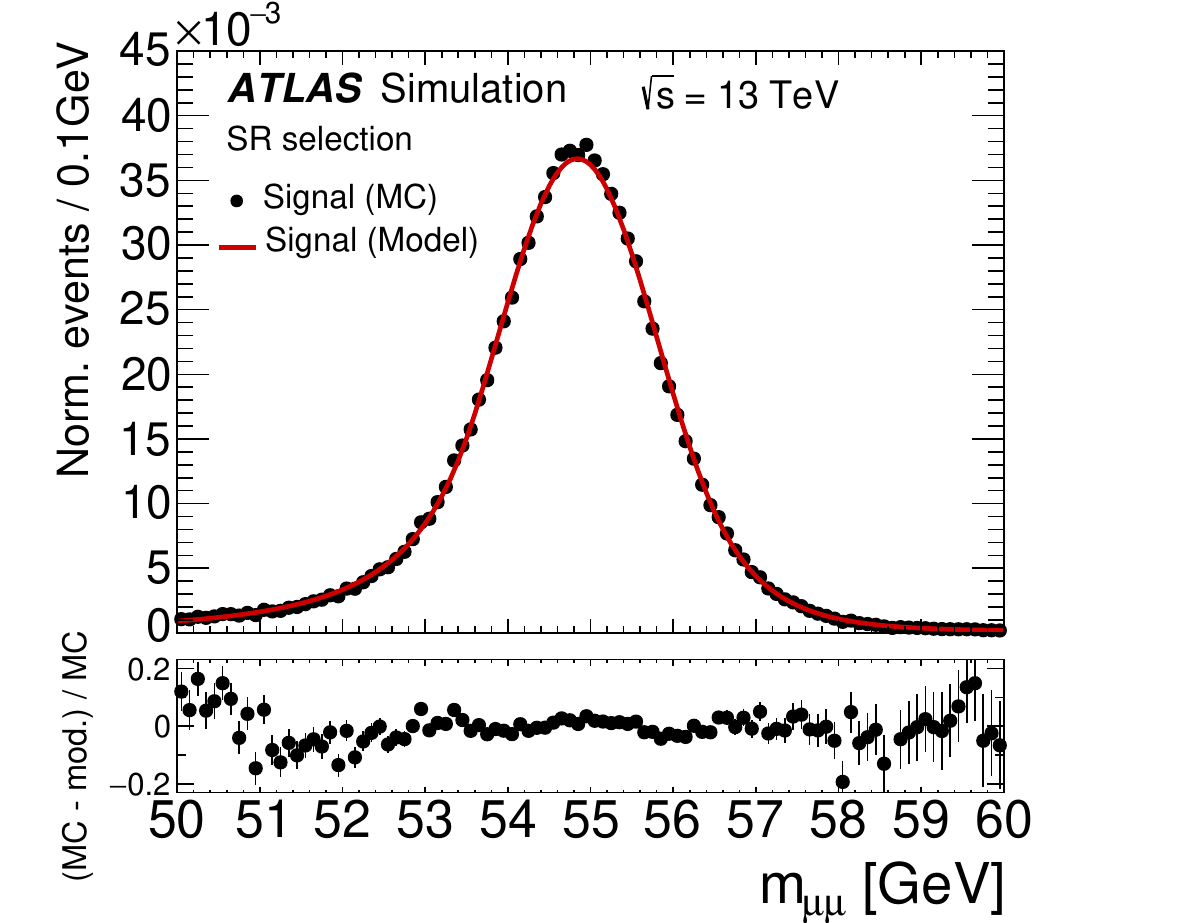} \label{fig:signalshape}}
\caption{(a) Width of the DSCB Gaussian core as a function of the new resonance mass (solid markers), with a linear fit overlaid. (b) Simulated dimuon mass distribution for a hypothetical $Z^\prime$ resonance with a mass of 55~\GeV and a width $\Gamma$ = 0.21~\GeV (solid markers), overlaid with the signal model, defined as a Breit–Wigner distribution convolved with a DSCB function and obtained from a parametric fit to the signal template. The lower panel displays the relative difference between the MC simulation and the signal model.}
\end{figure}

The parameters of the DSCB function are extracted by fitting simulated signal samples. The width of the Gaussian core ($\sigma$), which reflects the detector resolution, varies from approximately 0.6~\GeV to 1.3~\GeV, depending on the new resonance mass. To enable a smooth scan over the full mass range from 35~\GeV to 75~\GeV, a continuous signal model is constructed by interpolating the fitted parameters obtained from a set of discrete mass point simulations, as shown in Figure~\ref{fig:resolution}. Figure~\ref{fig:signalshape} presents a representative comparison between the simulated dimuon mass distribution and the parameterised signal shape, demonstrating the accuracy of the signal modelling.

\section{Systematic uncertainties}

Systematic uncertainties affecting both the fitted signal yield and shape are considered. These include theoretical, experimental, and background-modelling uncertainties. The latter play a dominant role in the analysis and are discussed in detail in Section~\ref{subsec:hypertuning}.

To estimate theoretical uncertainties in the signal yield, systematic variations are applied to account for uncertainties in the choice of QCD renormalisation ($\mu_\text{R}$) and factorisation ($\mu_\text{F}$) scales, as well as in the PDFs. The QCD scales are varied independently by factors of 0.5 and 2.0 relative to their nominal values, with the largest observed deviation taken as the systematic uncertainty. Following PDF4LHC recommendations~\cite{Butterworth:2015oua}, PDF uncertainties are estimated by using the envelope of variations across different PDF sets.

As discussed in Section~\ref{sec:samples}, the signal cross-section is normalised to the NNLO prediction, and uncertainties from Ref.~\cite{AH:2020cok} are used to estimate cross-section uncertainties. The NNLO cross-section uncertainty, together with the effects of QCD scale and PDF variations on the acceptance, is considered only when interpreting the cross-section limits as constraints on the coupling parameters.

Experimental systematic uncertainties related to the reconstructed physics objects also affect expected yields. In addition, uncertainties in the muon momentum scale and resolution impact the signal mass distribution. The experimental uncertainties considered include muon reconstruction and identification efficiencies, efficiencies related to the trigger, isolation and impact-parameter requirements, and the muon momentum scale and resolution~\cite{MUON-2018-03,PERF-2015-10,TRIG-2018-01}. The uncertainty in the combined 2015–2018 integrated luminosity is 0.83\%~\cite{DAPR-2021-01}, derived using the LUCID-2 detector~\cite{Avoni_2018} for the primary luminosity measurements. The overall relative experimental uncertainty in the dimuon event selection efficiency is estimated to be below 1\%, while the total uncertainty in the dimuon mass resolution is approximately 3.5\%.

All systematic uncertainties impacting the final dimuon invariant mass distribution are implemented as nuisance parameters, constrained by Gaussian likelihood terms in the statistical analysis described in Section~\ref{subsec:SBGP}.

% End of text imported from the .//sections/signalmodelling.tex input file

% The next lines are included from the .//sections/backgroundmodelling.tex input file
\section{Background modelling and signal yield extraction}
\label{sec:backgroundmodelling}

\begin{figure}[!htbp]
\centering
\includegraphics[width=0.49\textwidth,page=1]{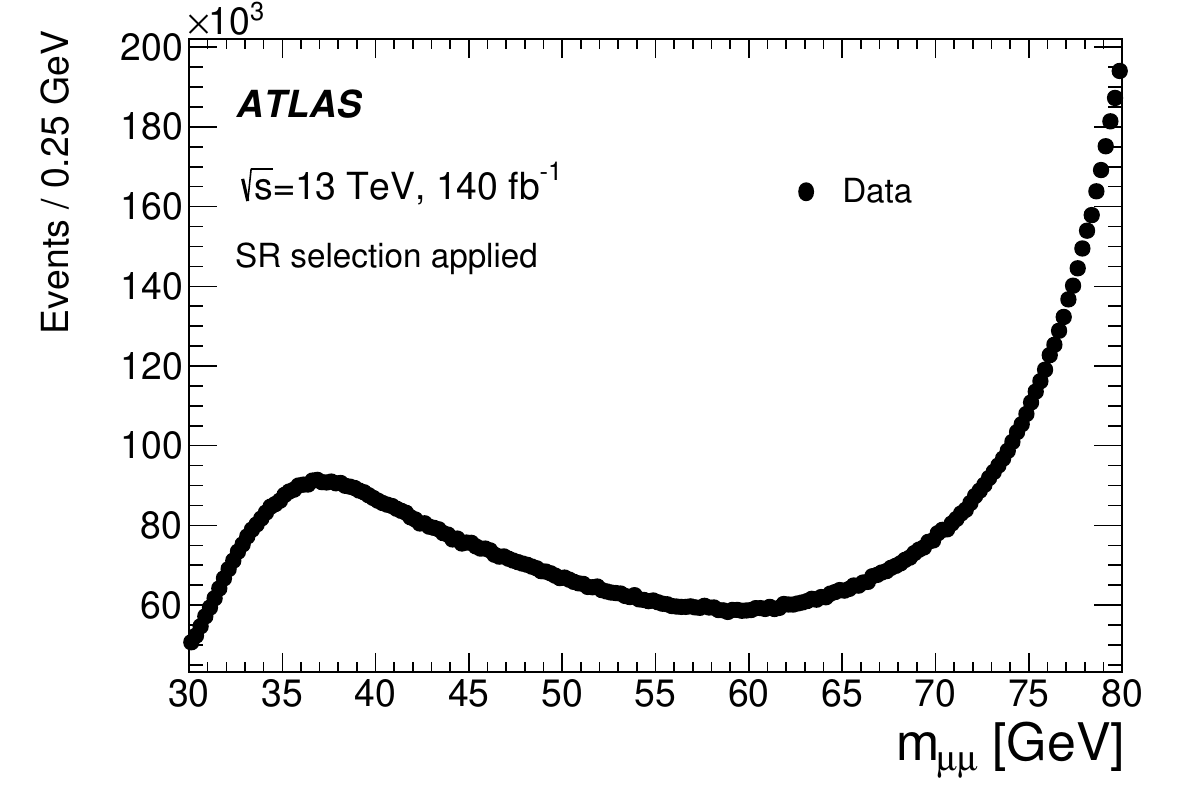}
\caption{The dimuon mass, \(m_{\mu\mu}\), distribution in the range 30--80~\GeV, after applying the selections described in Section~\ref{sec:eventselection}.}
\label{fig:dimuon_lowmass}
\end{figure}

The analysis adopts a bump-hunting strategy, fitting the mass spectrum using a data-driven estimation of the background. However, as shown in Figure~\ref{fig:dimuon_lowmass}, the mass spectrum exhibits a non-trivial structure, including threshold effects from the muon triggers that are most visible in the 30--45~\GeV region. This complexity makes it difficult to model the background using traditional analytic functions typically used for smoothly falling spectra.

To overcome these challenges, GPR is employed. GPR, a well-established machine-learning technique typically used for smoothing procedures~\cite{HIGG-2019-23}, provides a flexible yet robust framework for background modelling. In this analysis, it is used innovatively to model the background within a simultaneous signal-plus-background (S+B) fit, enabling signal extraction.

The GPR assumes that bin-to-bin correlations in the background \(m_{\mu\mu}\) spectrum follow a multivariate Gaussian distribution. Under this assumption, the background distribution is modelled as a Gaussian random process, with the covariance matrix determining the degree of correlation between bins.

The flexibility of the GP is governed by the covariance matrix, derived from the chosen kernel function, and the prior mean function. The radial basis function (RBF) kernel is selected in this analysis, with the correlation decreasing as a function of the distance between bins:
\begin{equation}
\label{eq:RBF}
C_{ij} = c(x_{i}, x_{j}) = k \exp \left(-\frac{(x_i-x_j)^2}{\ell^2} \right),
\end{equation}
where $k$ and $\ell$ are the covariance strength and length scale, respectively. The tuning of these hyper-parameters, aimed at optimising sensitivity and minimising bias, is detailed in Section~\ref{subsec:hypertuning}. Alternative kernel choices were explored during the development of the GPR. More complex kernels, such as the Gibbs kernel~\cite{remes2017nonstationaryspectralkernels}, introduce additional hyperparameters that are less straightforward to interpret and optimise. In the studies conducted, these kernels did not lead to improvements in statistical uncertainty or a reduction in the background modelling bias compared to the RBF kernel. The RBF kernel is therefore adopted due to its simplicity, interpretability, and robust performance. The prior mean is parameterised by a ninth-degree polynomial function fitted to the generator-smeared samples. As the prior mean is determined before any reweighting of these samples, as described at the end of Section~\ref{subsec:bkgsamples}, the resulting bias in the uncertainty estimation from background modelling (Section~\ref{subsec:hypertuning}) is negligible and does not artificially improve the agreement between the background model and the data. It has also been checked that modest variations of the polynomial parameterisation do not lead to any significant change in the results.

\subsection{Signal extraction procedure with GPR-based background modelling}
\label{subsec:SBGP}

The GPR is applied to the \(m_{\mu\mu}\) distribution, binned as $(m_i, N_i)$ for $i=1,\ldots,n$, where $m_i$ is the bin centre and $N_i$ is the observed yield. Signal extraction is incorporated into the fit through the following likelihood:
\begin{align*}
L(\text{data} | \mu, \theta) ={}&
\left( \prod_{i\in \mathrm{bins}} G\!\left(N_{i}^{\mathrm{background}}(\mu, \theta) \,\big|\, N_{i}^{\mathrm{GP}}, \sqrt{N_{i}^{\mathrm{GP}}}\right) \right) \nonumber \\
&\times
\left( G(\bm{N_{\mathrm{GP}}} \mid \bm{b_{0}}, \bm{C_{0}}) \right) \nonumber \\
&\times
\left( \prod_{i \in \mathrm{systematics}} G(\theta_{i} \mid \theta_{i}^{\mathrm{aux}}, \sigma(\theta_{i}^{\mathrm{aux}}))  \right)
\end{align*}

where the background yield in bin $i$ is given by
\begin{equation*}
N_i^{\text{background}} = N_i^{\text{data}} - N_i^{\text{signal}}(\mu, m_X, \Gamma, \theta)
\end{equation*}
Here, $N_i^{\text{data}}$ is the observed data yield in bin $i$, $N_i^{\text{signal}}(\mu, m_X, \Gamma, \theta)$ denotes the expected signal yield for the given signal hypothesis and $\mu$ is the signal strength, relative to a reference signal yield. The first term in the likelihood compares the GP background model ($N_{i}^{\text{GP}}$) to the observed background ($N_i^{\text{background}}$) across different bins. The second term, introduced by the GPR, constrains the background prediction to the prior mean function ($\bm{b_{0}}$) through the covariance matrix ($\bm{C_0}$) determined by the kernel function in Eq.~(\ref{eq:RBF}), where $\mathbf{N}_{\mathrm{GP}}$ denotes the binned vector of the predicted yields from GP. The third term applies Gaussian constraints to the nuisance parameters $\theta_i$, each associated with an auxiliary measurement $\theta_i^{\mathrm{aux}}$ and its corresponding uncertainty $\sigma(\theta_i^{\mathrm{aux}})$. These constraints incorporate systematic uncertainties into the likelihood and enable a simultaneous fit to signal and background within a data-driven framework, ensuring a statistically well-defined treatment of all sources of uncertainty. Based on this likelihood, the best-fit mean and covariance functions, derived from the GP using the input histogram, define the background shape.

The parameter of interest extracted from the likelihood fit is the signal strength $\mu$, which serves as a scaling factor for the total event yield predicted by the signal model. This is subsequently converted into the fiducial cross-section using a correction factor for the detection efficiency.

\subsection{Hyper-parameters tuning of GPR}
\label{subsec:hypertuning}
The GP kernel function captures the correlation structure across bins in the likelihood fit. The choice of kernel hyper-parameters is critical, as it controls the flexibility and smoothness of the background modelling. In the RBF kernel used in this analysis, the hyper-parameter $k$ defines the correlation strength across bins, while $\ell$ characterises the correlation length scale, i.e., how far neighbouring bins influence each other.

The choice of hyper-parameters is guided by their ability to model the background shape reliably while maintaining a balance between flexibility and robustness. The values of $k$ and $\ell$ are selected to achieve a smooth yet accurate background model, following the procedure outlined below. The optimisation is performed on mass-spectrum templates reweighted from the generator-smearing sample to the fully simulated sample, as described in Section~\ref{subsec:bkgsamples}. The criteria for choosing suitable hyper-parameters include the requirement that only hyper-parameter sets yielding a $\chi^2$ probability greater than 1\% when fitting the simulation template are retained. For the final choice of parameters, the background-only fit $\chi^2$ probability is in the range of 5--20\% when evaluated in a \(\pm3\sigma\) mass window around each tested mass point, where \(\sigma\) is the detector resolution at that mass.

To validate the signal extraction procedure, a series of signal injection tests is performed. These include scenarios with varying injected signal strengths as well as a null injection case.  The response to a null signal injection, referred to as a spurious signal (SS)~\cite{HIGG-2012-27}, is evaluated by fitting signal-plus-background models to background-only templates. Any non-zero fitted signal yield in this context indicates an intrinsic bias in the background modelling in the absence of a true signal. The magnitude of the SS is treated as a systematic uncertainty associated with the background model. Following recommendations for smooth background modelling~\cite{ATL-PHYS-PUB-2020-028}, the background model is optimised to ensure that the SS yield satisfies
$N_\text{SS} < 0.5\sigma_{\text{fit}}$, where $\sigma_{\text{fit}}$ is the statistical uncertainty in the fitted signal yield. This criterion ensures that any potential bias remains within the range covered by the modelling uncertainty controlled via nuisance parameters.

Additional validation tests are conducted by injecting signals of varying strengths on top of the background-only template. The linearity between the injected and extracted signal yields is verified by fitting the response curve, and the fitted slope is required to remain within 3\% of unity during hyper-parameter optimisation. Any residual deviation is treated as an additional systematic uncertainty associated with the signal extraction procedure.

Among the hyper-parameter configurations meeting the established criteria, the final choice is made by selecting the set that minimises the total uncertainty in the extracted signal yield. This total uncertainty is defined as the quadrature sum of the statistical uncertainty from the fit and the systematic contribution from the SS, thereby ensuring optimal overall sensitivity.

\begin{figure}[!htbp]
\centering
\includegraphics[width=0.5\textwidth]{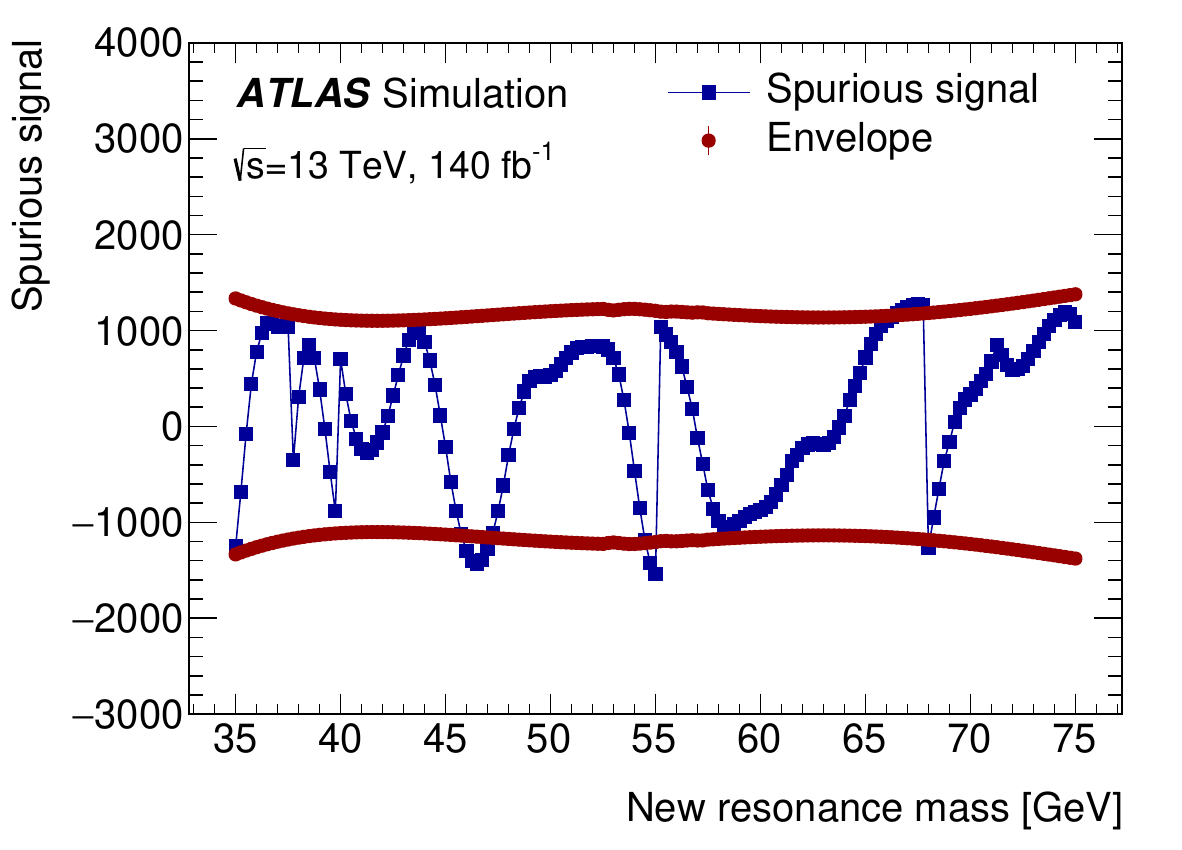}
\caption{ Spurious signal yield as a function of the new resonance mass, obtained by fitting signal-plus-background models to background-only templates. The points represent the fitted signal yields, with vertical error bars indicating the corresponding statistical uncertainties. The thick curves denote the symmetrised envelope, which defines the systematic uncertainty associated with the background modelling. Discontinuities in the spurious signal reflect transitions between different Gaussian process hyper-parameter configurations.}
\label{fig:dimuon_lowmass_sim}
\end{figure}

Different hyper-parameter configurations are used for distinct resonance mass ranges, as summarised in Table~\ref{tab:gpconfiguration}.
The optimal values depend on the resonance mass, reflecting variations in the background shape and signal resolution; consequently, no single set of hyper-parameters provides an adequate description across the entire mass range.
The parameter $k$ correlates with the behaviour of the spectrum shown in Figure~\ref{fig:dimuon_lowmass}. Larger
$k$ values are used near the rapidly varying edges of the mass range to constrain the GP prediction around the prior mean, while smaller values in the central region allow greater flexibility.
The correlation length $\ell$ is primarily driven by the signal resolution and generally increases with the resonance mass, except in the few~\GeV\ at the upper edge of the mass range, where smaller $\ell$ values are required to capture the steeply varying spectrum.
The resulting SS values after optimisation are shown in Figure~\ref{fig:dimuon_lowmass_sim} as a function of the resonance mass.
The envelope of the SS distribution is obtained as the magnitude of the analytic signal constructed from the quadrature sum of the original signal and its Hilbert-transformed~\cite{Cohen1995} counterpart, and is subsequently smoothed with a Savitzky--Golay filter. This envelope defines the final systematic uncertainty associated with the background model, and its symmetrised band is taken as the systematic uncertainty in the background modelling. The sensitivity of the analysis is dominated by statistical uncertainties, while the leading systematic contribution arises from the spurious signal uncertainty, with a relative impact of approximately 20\%.

\begin{table}[htbp]
\centering
\caption{RBF kernel hyper-parameter sets optimised for different $m_X$ mass hypotheses, tuned using spurious signal and linearity tests.}
\begin{tabular}{r@{\,$\leq$\,}c@{\,$<$\,}l | c | c}
\hline
\multicolumn{3}{c|}{$m_{X}$ [GeV] range} & $k$ & $\ell$ [GeV] \\
\hline
35   & $m_X$ & 38   & $4\times10^{9}$ & 4.4  \\
38   & $m_X$ & 40   & $2\times10^{8}$ & 8.0  \\
40   & $m_X$ & 55   & $2\times10^{8}$ & 7.1  \\
55   & $m_X$ & 68   & $2\times10^{8}$ & 8.7  \\
68   & $m_X$ & 71.5 & $5\times10^{8}$ & 12.1 \\
71.5 & $m_X$ & 75   & $5\times10^{8}$ & 11.4 \\
\hline
\end{tabular}
\label{tab:gpconfiguration}
\end{table}

% End of text imported from the .//sections/backgroundmodelling.tex input file

% The next lines are included from the .//sections/statistical.tex input file
\section{Statistical interpretation and limit setting}
\label{sec:ts}
Hypothesis testing of the signal strength $\mu$ is based on the profile likelihood ratio~\cite{Cowan:2010js}, with the test statistic for the limit-setting:
\begin{equation}
q_\mu =
\begin{cases}
-2\ln \frac{L(\text{data} \mid \mu, \hat{\hat{\theta}}(\mu))}{L(\text{data} \mid 0, \hat{\hat{\theta}}(0))}, & \hat{\mu} < 0\\
-2\ln \frac{L(\text{data} \mid \mu, \hat{\hat{\theta}}(\mu))}{L(\text{data} \mid \hat{\mu}, \hat{\theta})}, & 0 \leq \hat{\mu} \leq \mu \\
0. & \hat{\mu} > \mu
\end{cases}
\label{eq:qu}
\end{equation}
Here, $\hat{\mu}$ and $\hat{\theta}$ denote the unconditional
maximum-likelihood (ML) estimators of the signal strength and nuisance
parameters, while $\hat{\hat{\theta}}(\mu)$ and $\hat{\hat{\theta}}(0)$ are
the corresponding conditional estimators obtained under the signal-plus-background and background-only hypotheses, respectively.
The likelihood function $L$ incorporates all sources of systematic uncertainty
through nuisance parameters and their correlations. For significance and $p$-value evaluations, the corresponding one-sided test statistic $q_0$ is used, following the asymptotic formalism in Ref.~\cite{Cowan:2010js}.

The significance of a potential excess is first quantified with the $p_0$ value, defined as the probability that the background could fluctuate to the observed excess:
\begin{equation*}
p_0 = P(q_0 > q_0^{\text{obs}} \hspace{2mm}|\hspace{2mm} \mu = 0)
\end{equation*}

For discovery tests, the one-sided statistic $q_0$ is defined analogously to Eq. (\ref{eq:qu}), restricted to positive signal-strength estimates ($\hat{\mu} > 0$)~\cite{Cowan:2010js}. The $p_0$ value is converted into a local significance, expressed as the equivalent number of Gaussian standard deviations ($\sigma$) assuming a one-sided normal distribution.

Upper limits on the signal strength are evaluated using the CL\(_\text{s}\) method~\cite{Read:2002hq}. A signal strength \(\mu\) is considered excluded at the 95\% confidence level when the resulting CL\(_\text{s}\) value falls below 0.05.

Since the measurement is performed within a fiducial phase-space region (defined in Section~\ref{sec:eventselection}), to facilitate reinterpretation, the number of signal events $N_\text{S}$ is parameterised in terms of the fiducial cross-section times branching ratio to derive upper limits:
\begin{equation*}
\sigma_{\text{fid}} \times \mathcal{B}(X\to\mu\mu) = \frac{N_\text{S}}{\varepsilon \times \mathcal{L}},
\end{equation*}
where the $\mathcal{L}$ is the integrated luminosity and $\varepsilon$ is the correction factor accounting for the detector efficiency, described in Section~\ref{sec:signalmodelling}.

% End of text imported from the .//sections/statistical.tex input file

% The next lines are included from the .//sections/result.tex input file
\section{Results}
\label{sec:results}

\begin{figure}
\centering
\subfloat[]{\includegraphics[width=0.5\textwidth]{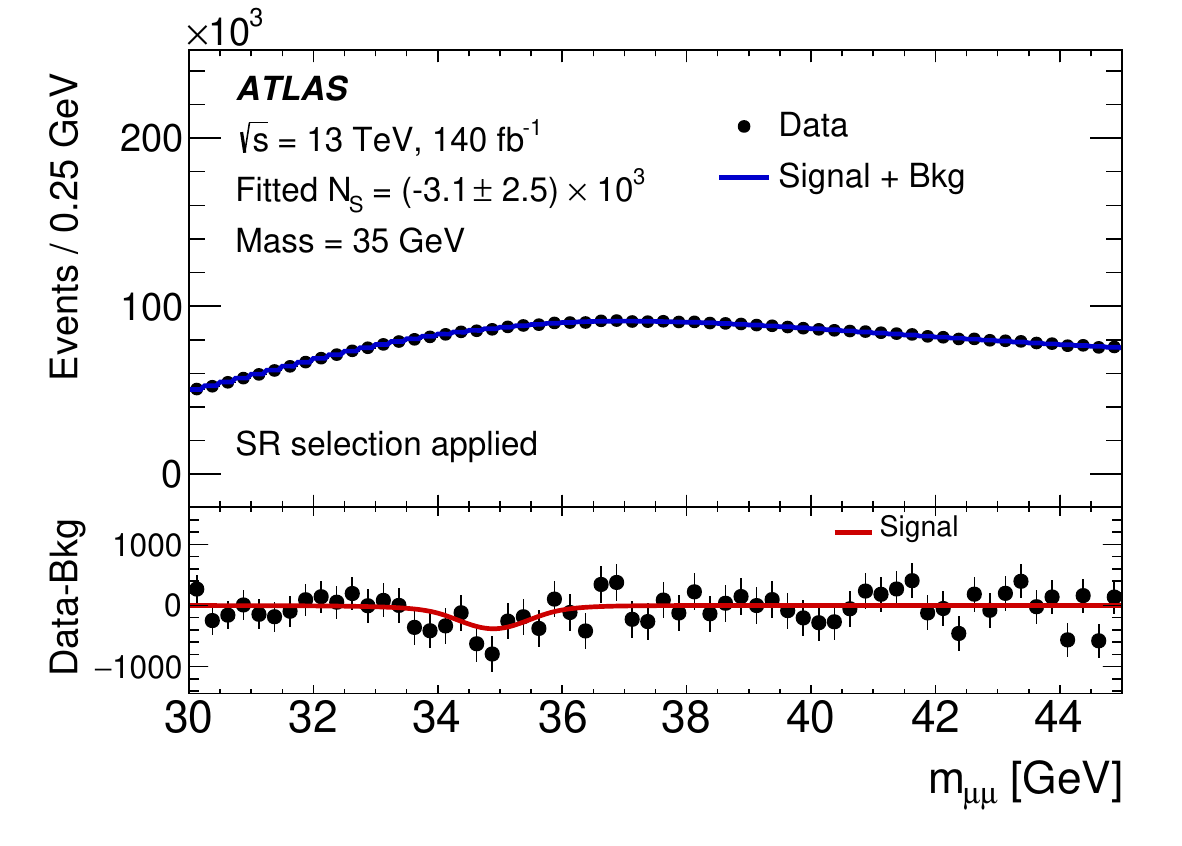}}
\subfloat[]{\includegraphics[width=0.5\textwidth]{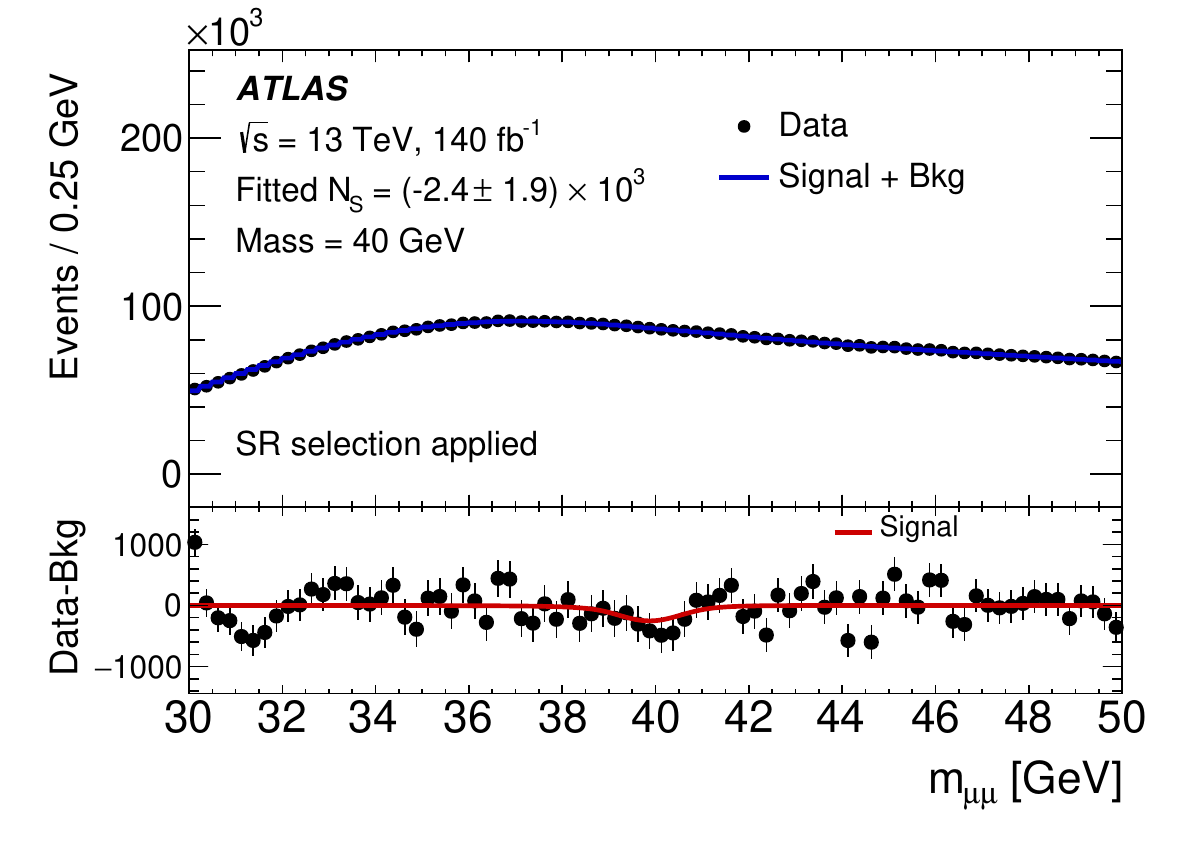}}\\
\subfloat[]{\includegraphics[width=0.5\textwidth]{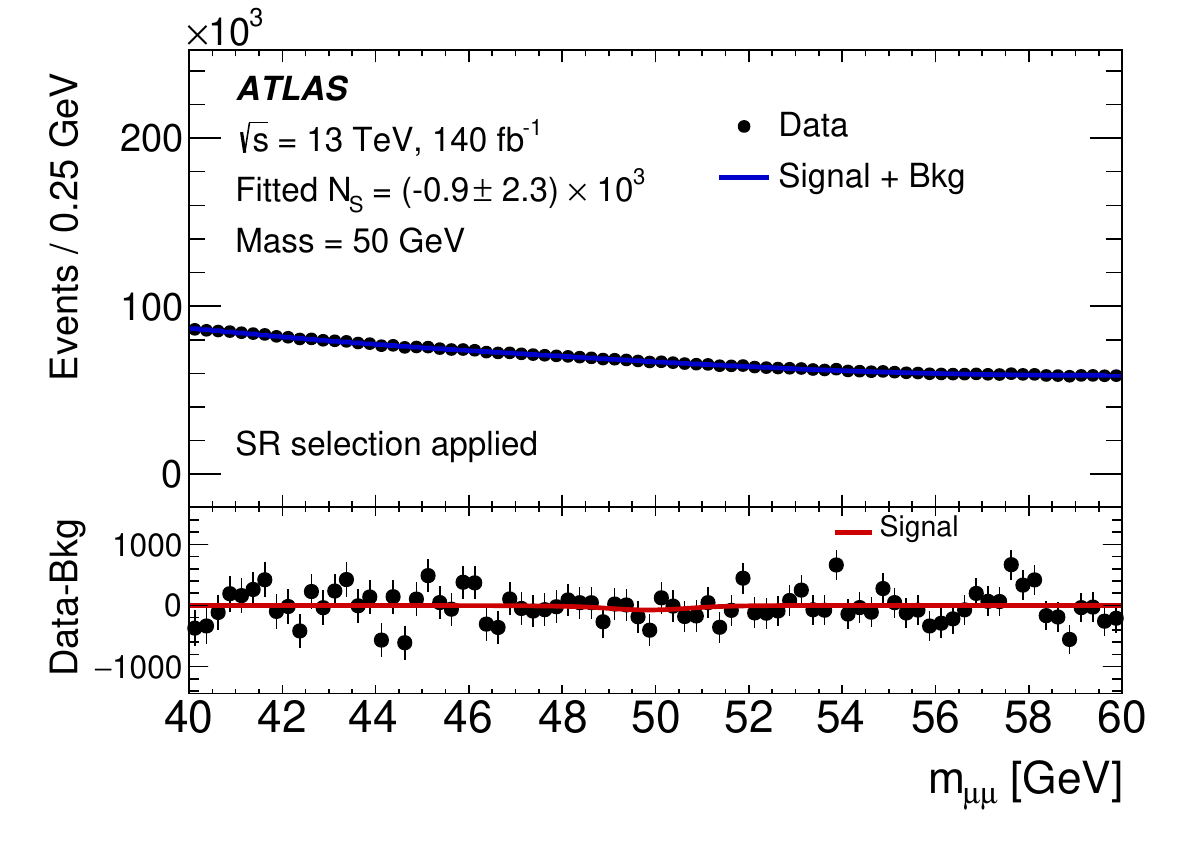}}
\subfloat[]{\includegraphics[width=0.5\textwidth]{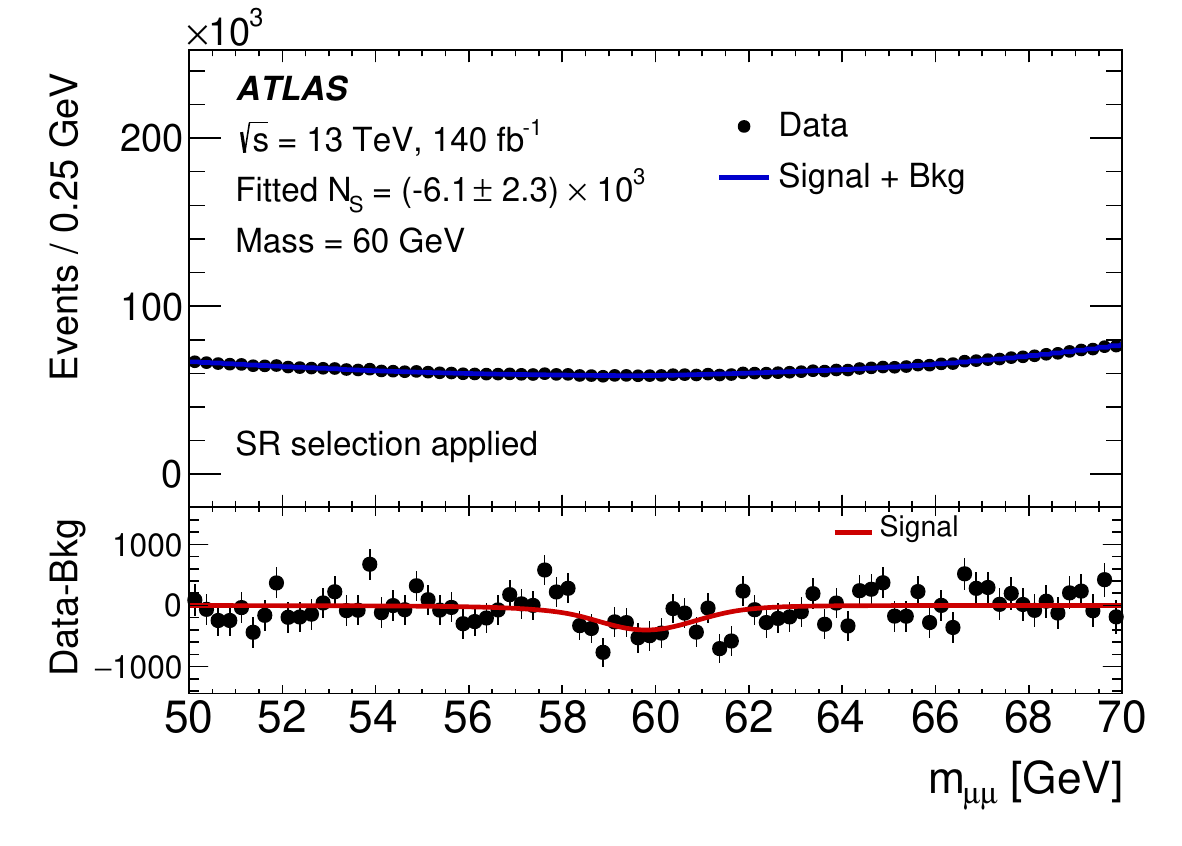}} \\
\subfloat[]{\includegraphics[width=0.5\textwidth]{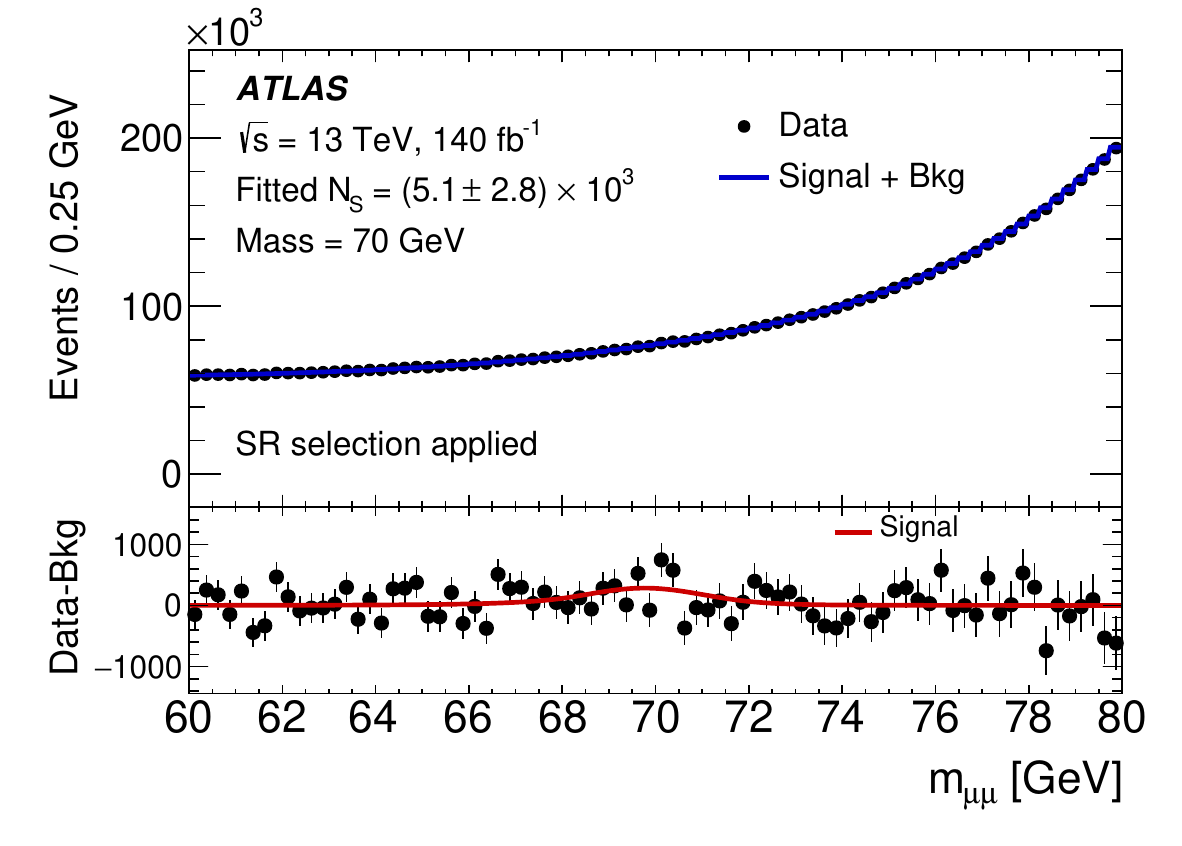}}
\subfloat[]{\includegraphics[width=0.5\textwidth]{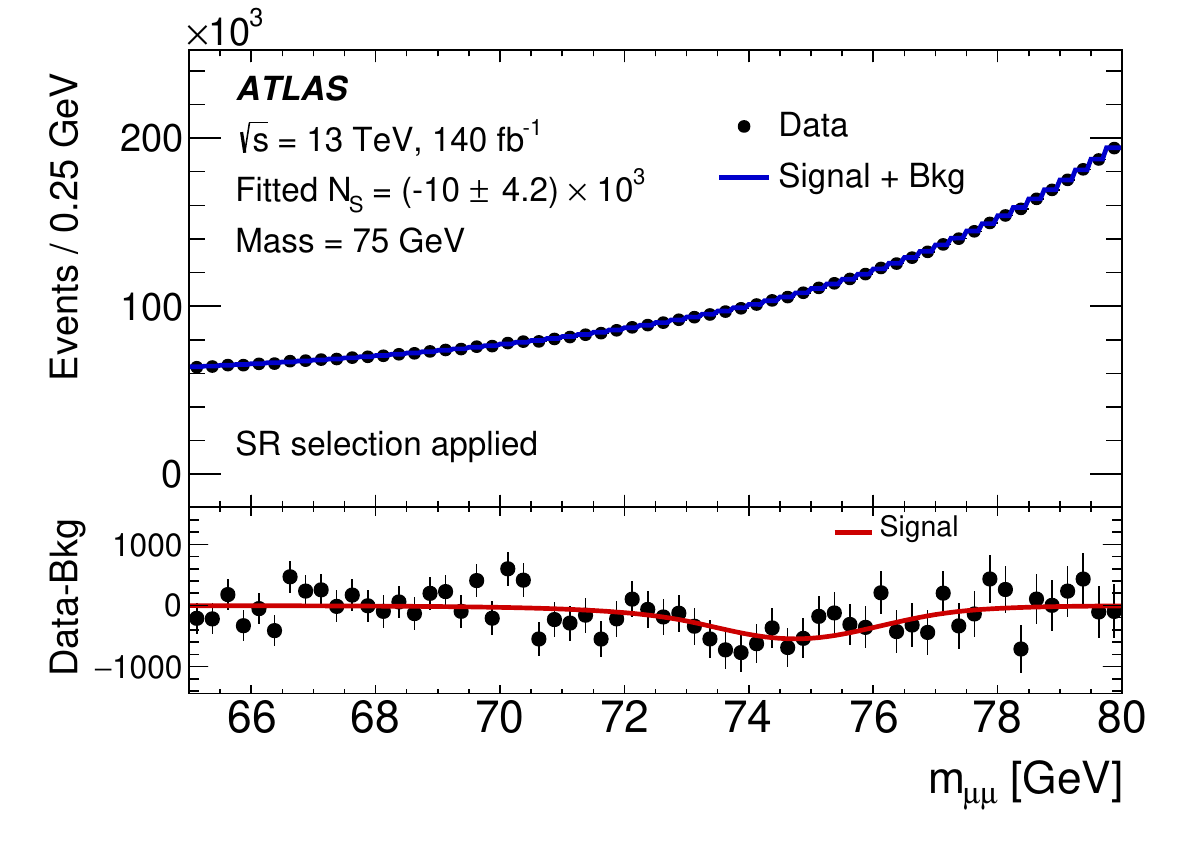}}\\

\caption{Fit results for various new resonance mass hypotheses. Each plot shows the \(\text{m}_{\mu\mu}\) distribution in data overlaid with the fitted signal-plus-background model. The lower panels display the data-minus-background residuals (points), obtained by subtracting the estimated background component of the fit model from the data, and the extracted signal component (line). The $\text{N}_\text{S}$ in the plots refers to the extracted signal yield. The fits are performed in the range of 30~\GeV to 80~\GeV, but only the region around the signal hypothesis is shown. The plots correspond to different resonance mass hypotheses: (a) 35~\GeV, (b) 40~\GeV, (c) 50~\GeV, (d) 60~\GeV, (e) 70~\GeV, and (f) 75~\GeV.}
\label{fig:massfit}
\end{figure}

The signal extraction is performed by fitting the dimuon invariant mass spectrum in the range of 30--80~\GeV, which corresponds to events satisfying the signal-region (SR) selection defined in Section~\ref{sec:eventselection}. The search for a potential resonance is carried out within the narrower range of 35--75~\GeV to avoid potential edge effects from the fit boundaries. The dimuon mass distributions, overlaid with signal-plus-background fits using various GP hyperparameter configurations, are shown in Figure~\ref{fig:massfit} for several representative resonance mass hypotheses.

\begin{figure}[htb]
\centering
\includegraphics[width=0.8\linewidth]{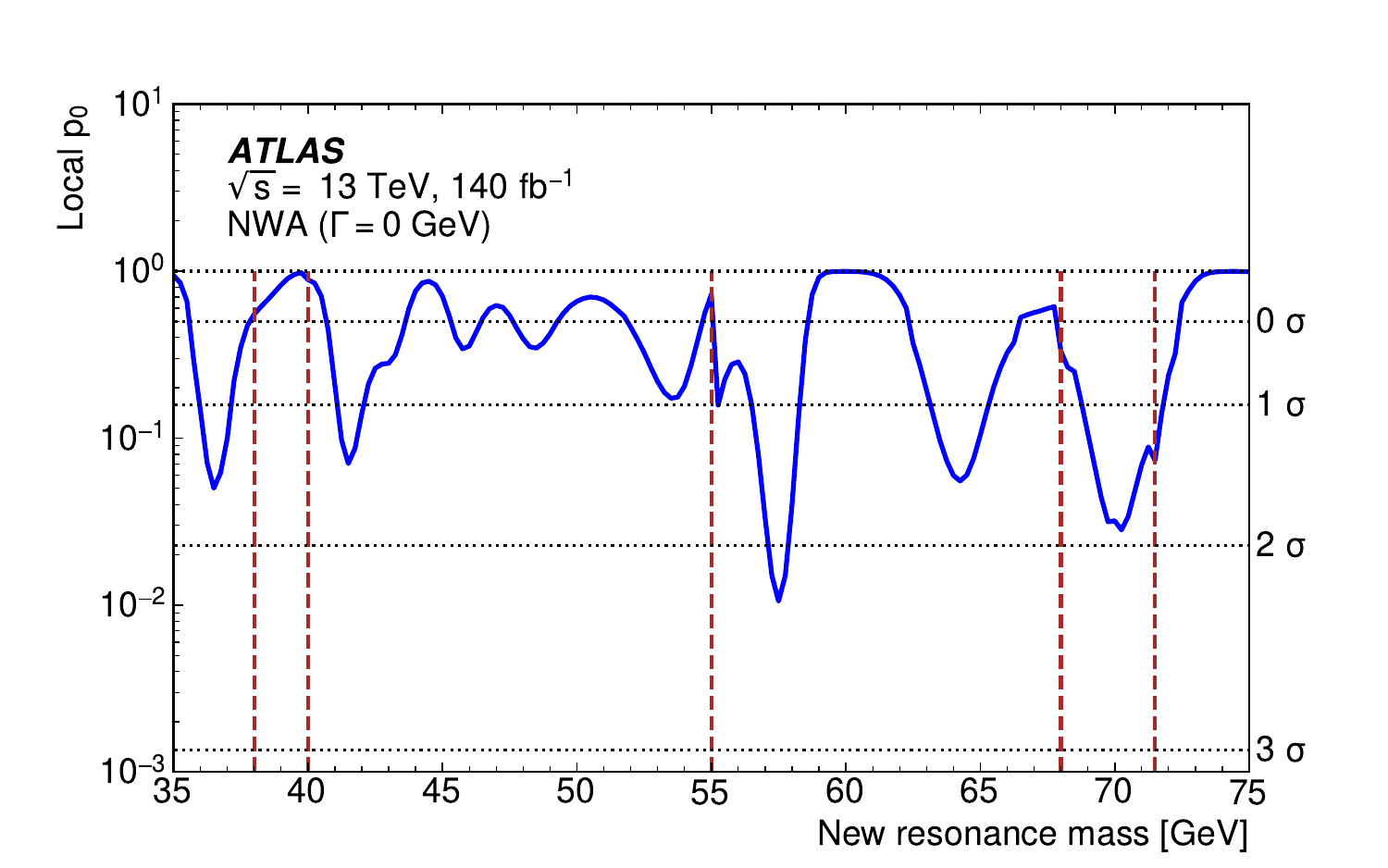}
\caption{Scan of the $p_0$ values as a function of the new resonance mass. Discontinuities correspond to the boundaries between different Gaussian process hyper-parameter regions. The vertical dashed lines indicate the mass points where the hyper-parameters are changed.}
\label{fig:pvalue}
\end{figure}

The $p_0$ values and local significance, corresponding to the background-only hypothesis, are scanned across a range of hypothesised new resonance masses, as shown in Figure~\ref{fig:pvalue}. The methodology for this calculation is detailed in Section~\ref{sec:ts}. The minimum $p_0$ value is observed at a mass of 57.5~\GeV, corresponding to a local significance of 2.3 $\sigma$, while the global significance is negligible. The ranking of systematic uncertainties is dominated by the SS contribution, followed by theoretical and experimental sources. %

\begin{figure}[htb]
\centering
\includegraphics[width=0.5\linewidth]{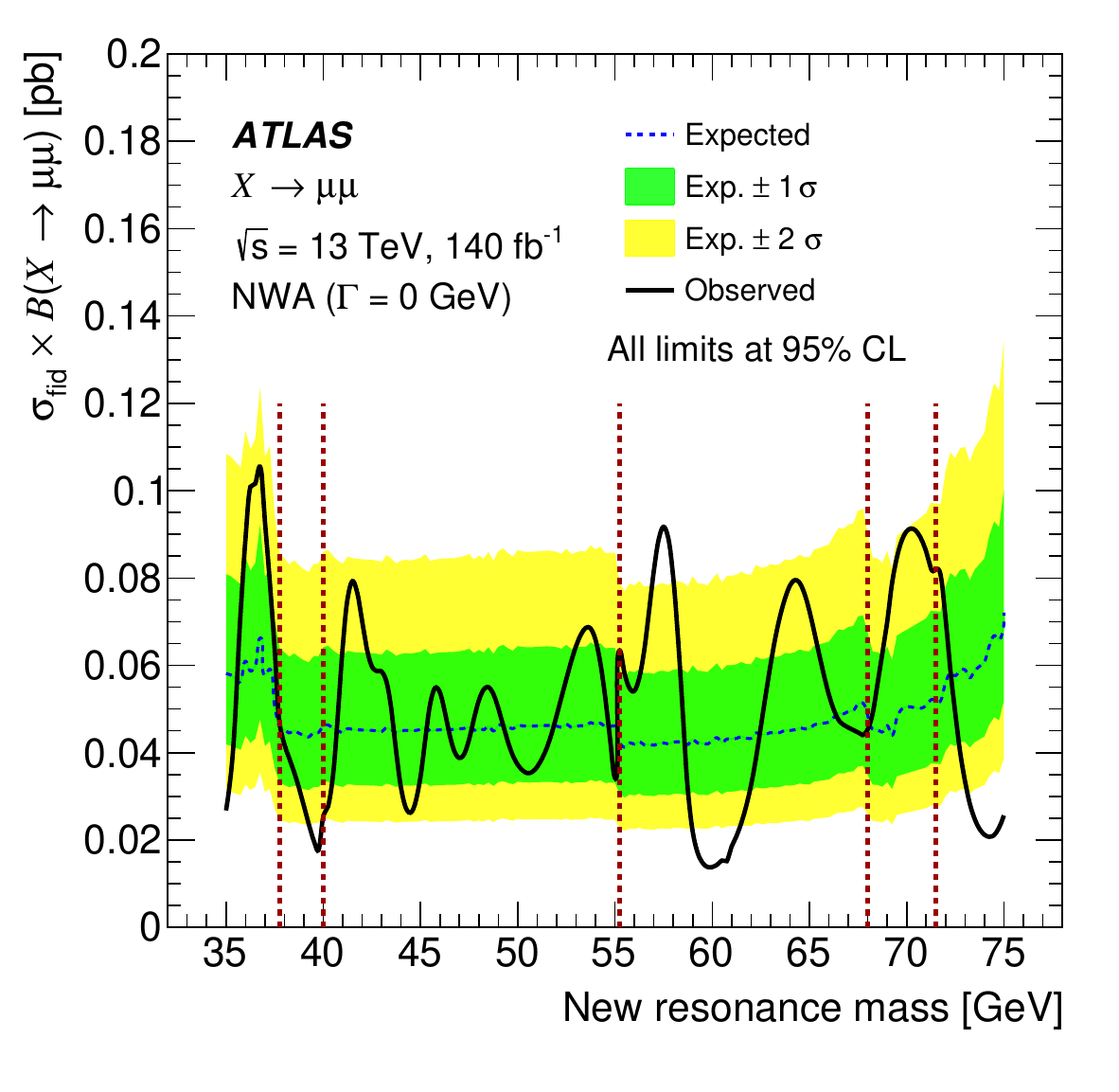}\caption{Observed and expected 95\% CL upper limits on the production fiducial cross-section times branching ratio of a narrow-width ($\Gamma$ = 0 GeV) resonance, as a function of the new resonance mass. Discontinuities correspond to the boundaries between different GP hyper-parameter regions. The vertical dashed lines indicate the mass points where the hyper-parameters are changed.}
\label{fig:upperlimit}
\end{figure}

The observed and expected 95\% confidence level (CL) upper limits on the production fiducial cross-section times branching ratio, $\sigma_{\text{fid}}\times \mathcal{B}(X\to\mu\mu)$, as a function of the resonance mass are shown in Figure~\ref{fig:upperlimit}. The limits are derived under the narrow-width approximation (\(\Gamma = 0\)). The results remain valid for resonances with relative intrinsic widths
\(\Gamma / m_X < 1\%\), covering a broad class of new-physics scenarios with perturbative couplings. The primary factors limiting the sensitivity are the size of the recorded data sample and uncertainties associated with the background modelling. Discontinuities in the limit curve reflect transitions between different sets of hyper-parameters used in the GP background modelling.

\begin{figure}[htb]
\centering
\subfloat[\label{fig:Zprimelimit}]{\includegraphics[width=0.49\linewidth]{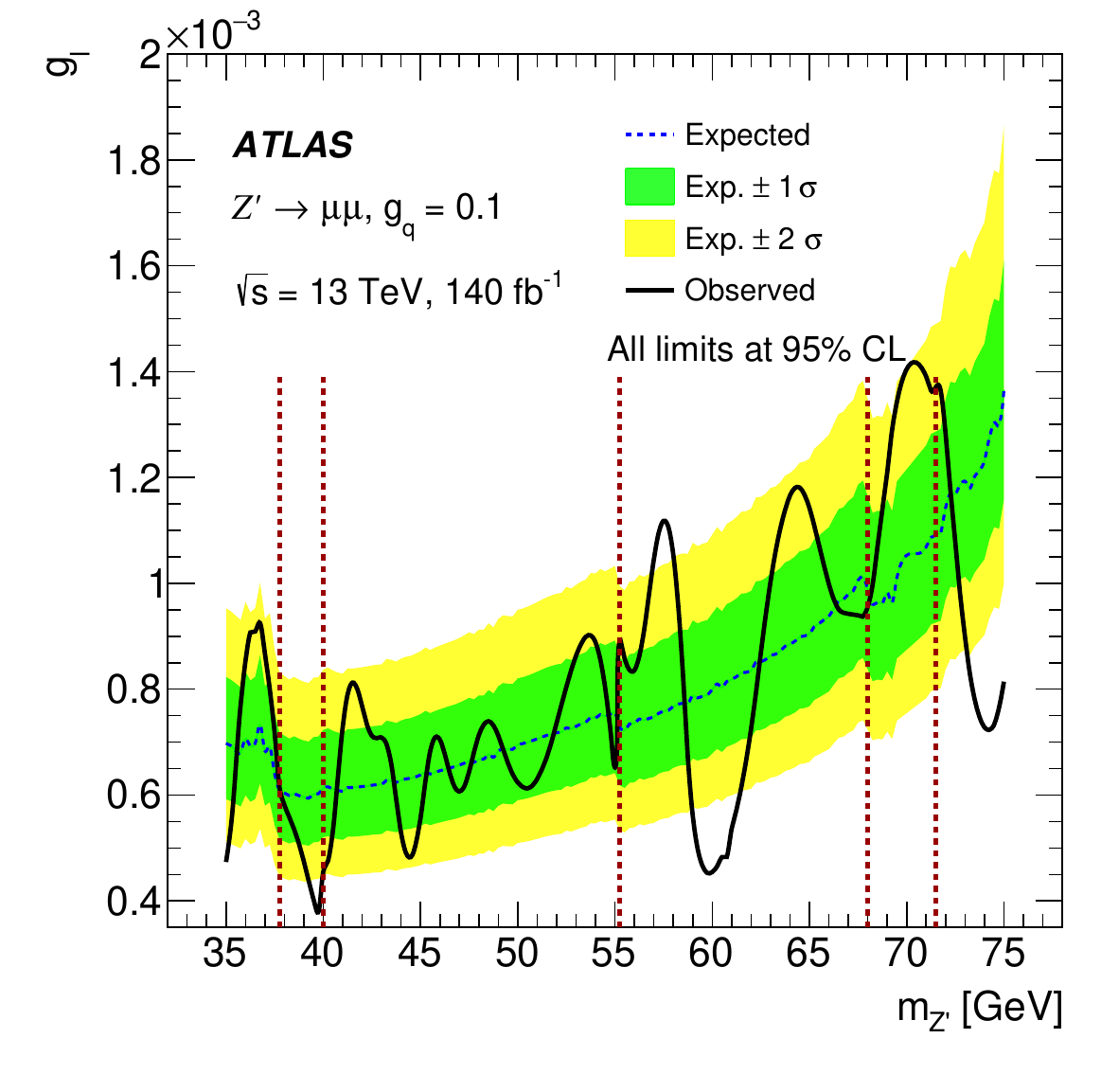}}
\subfloat[\label{fig:dplimit}]{\includegraphics[width=0.49\linewidth]{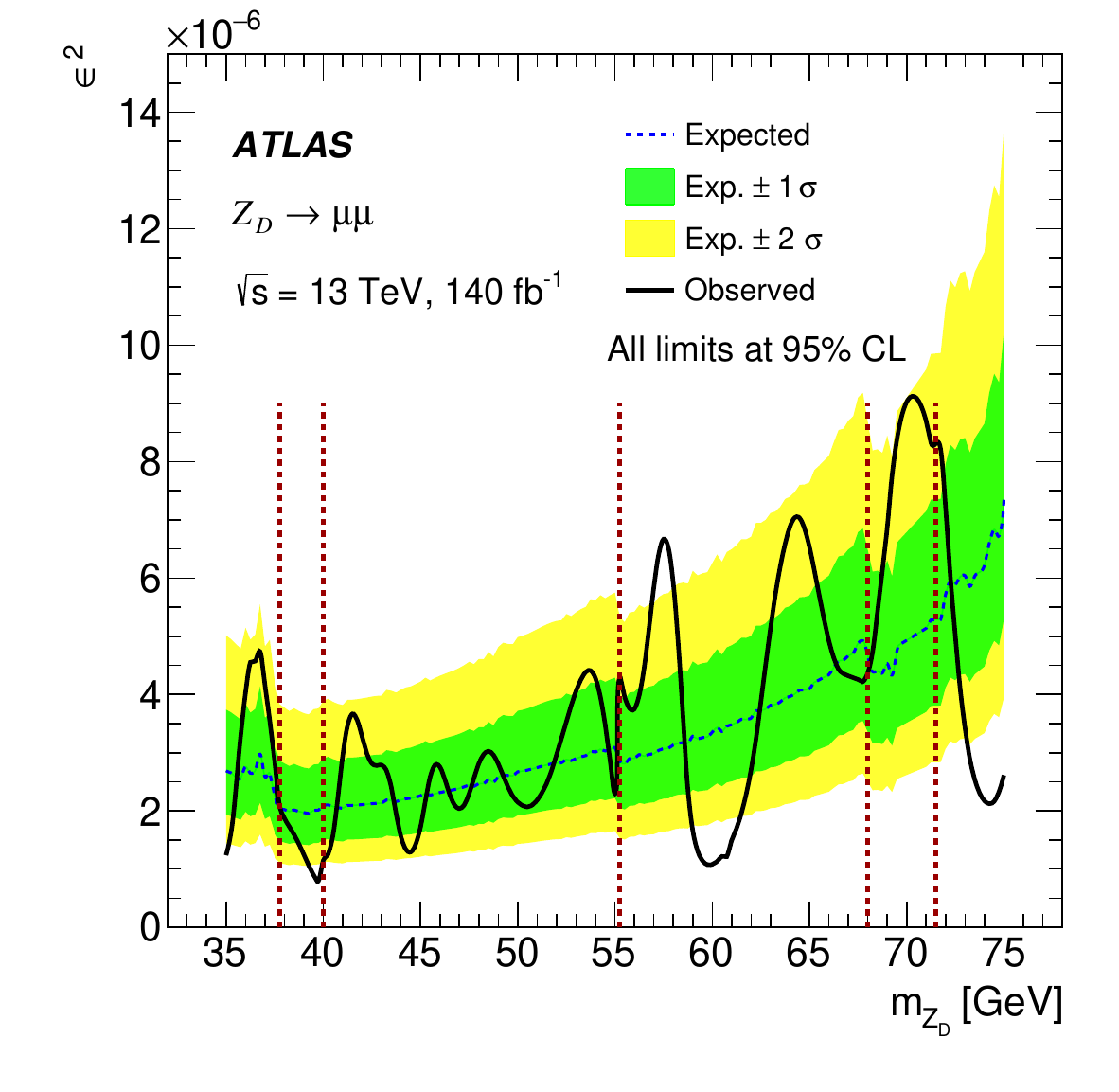}
}
\caption{(a) Observed and expected 95\% CL upper limits on the coupling between the $Z^\prime$ boson and leptons as a function of $m_{Z'}$, assuming $g_q=0.1$. (b) Observed and expected 95\% CL upper limits on the kinetic mixing parameter ($\epsilon^2$ in the dark photon model) as a function of $m_{Z_\text{D}}$. Discontinuities correspond to the boundaries between different GP hyper-parameter regions. Vertical dashed lines indicate the mass points where the hyper-parameters are changed.}
\label{fig:Zprime_signal_distribution2}
\end{figure}

Assuming a coupling of 0.1 between the $Z^\prime$ and quarks, an upper limit on the coupling ($g_l$) between $Z^\prime$ and leptons is derived by varying the intrinsic signal width as described in Section~\ref{sec:mc_samples}.
As shown in Figure~\ref{fig:Zprimelimit}, the limits on $g_l$ range from $4\times10^{-4}$ to $1.4\times10^{-3}$, depending on the assumed $Z^\prime$ mass.

A similar search strategy can be applied to models involving dark photons, where the coupling to SM particles is characterised by the kinetic mixing parameter $\epsilon$.
In such models, the production cross-section scales as $\epsilon^{2}$. The experimental sensitivity is thus expressed as an upper limit on $\epsilon^{2}$.
The resulting limits, shown in Figure~\ref{fig:dplimit}, range from $1\times10^{-6}$ to $9\times10^{-6}$.
As noted in Section~\ref{subsec:dimuonshapemodelling}, the narrow-width approximation is applicable to these models.

The search sets stringent constraints on the dark-photon kinetic mixing in the mass range 35–75~\GeV, complementing results from CMS~\cite{CMS-EXO-19-018} and LHCb~\cite{LHCb:2019vmc}. An improvement in sensitivity is achieved in the 35--45~\GeV region, where the CMS results~\cite{CMS-EXO-19-018} were obtained using a scouting data acquisition scheme. This improvement is driven by the mass-shape modelling capability provided by GPR, which enables the use of standard, unprescaled muon triggers together with the standard offline $p_\mathrm{T}$ selection, thereby suppressing backgrounds. In addition, the larger integrated luminosity of the dataset used in this analysis (140~fb$^{-1}$) compared to the CMS scouting dataset (96~fb$^{-1}$) also contributes to the improved sensitivity.

% End of text imported from the .//sections/result.tex input file

% The next lines are included from the .//sections/Conclusion.tex input file
\FloatBarrier
\section{Conclusions}
\label{sec:conclusions}

A search for new resonances in the dimuon invariant mass spectrum between 35~\GeV and 75~\GeV is performed using 140~$\text{fb}^{-1}$ of Run-2 proton–proton collision data at
$\sqrt{s}$ = 13~\TeV, collected with the ATLAS detector at the LHC\@. This study constitutes the first ATLAS search targeting this mass region, extending the sensitivity of previous dimuon resonance analyses to lower masses.

A novel background-modelling technique based on Gaussian process regression is employed, providing a flexible and accurate description of the rapidly varying Drell–Yan continuum and enhancing sensitivity to narrow resonances. The largest excess occurs near 57.5~\GeV, with a local significance of 2.3 standard deviations, while the corresponding global significance is negligible.

Upper limits are set on the fiducial production cross-section times branching ratio, defined within a fiducial region closely matching the detector-level selection. The observed limits range from 20~fb to 110~fb, depending on statistical fluctuations and variations in the background modelling configurations. The results are additionally interpreted in terms of constraints on theoretical parameters in both a simplified dark-matter mediator model and a dark-photon model.

% End of text imported from the .//sections/Conclusion.tex input file

%
%
%
%
%
\clearpage
\section*{Acknowledgements}
%

%

% The next lines are included from the .//acknowledgements/Acknowledgements.tex input file
%
%

%
%

We thank CERN for the very successful operation of the LHC and its injectors, as well as the support staff at
CERN and at our institutions worldwide without whom ATLAS could not be operated efficiently.

The crucial computing support from all WLCG partners is acknowledged gratefully, in particular from CERN, the ATLAS Tier-1 facilities at TRIUMF/SFU (Canada), NDGF (Denmark, Norway, Sweden), CC-IN2P3 (France), KIT/GridKA (Germany), INFN-CNAF (Italy), NL-T1 (Netherlands), PIC (Spain), RAL (UK) and BNL (USA), the Tier-2 facilities worldwide and large non-WLCG resource providers. Major contributors of computing resources are listed in Ref.~\cite{ATL-SOFT-PUB-2026-001}.

We gratefully acknowledge the support of ANPCyT, Argentina; YerPhI, Armenia; ARC, Australia; BMWFW and FWF, Austria; ANAS, Azerbaijan; CNPq and FAPESP, Brazil; NSERC, NRC and CFI, Canada; CERN; ANID, Chile; CAS, MOST and NSFC, China; Minciencias, Colombia; MEYS CR, Czech Republic; DNRF and DNSRC, Denmark; IN2P3-CNRS and CEA-DRF/IRFU, France; SRNSFG, Georgia; BMFTR, HGF and MPG, Germany; GSRI, Greece; RGC and Hong Kong SAR, China; ICHEP and Academy of Sciences and Humanities, Israel; INFN, Italy; MEXT and JSPS, Japan; CNRST, Morocco; NWO, Netherlands; RCN, Norway; MNiSW, Poland; FCT, Portugal; MNE/IFA, Romania; MSTDI, Serbia; MSSR, Slovakia; ARIS and MVZI, Slovenia; DSI/NRF, South Africa; MICIU/AEI, Spain; SRC and Wallenberg Foundation, Sweden; SERI, SNSF and Cantons of Bern and Geneva, Switzerland; NSTC, Taipei; TENMAK, T\"urkiye; STFC/UKRI, United Kingdom; DOE and NSF, United States of America.

Individual groups and members have received support from BCKDF, CANARIE, CRC and DRAC, Canada; CERN-CZ, FORTE and PRIMUS, Czech Republic; COST, ERC, ERDF, Horizon 2020 and Marie Sk{\l}odowska-Curie Actions, European Union; Investissements d'Avenir Labex, Investissements d'Avenir Idex and ANR, France; DFG and AvH Foundation, Germany; Herakleitos, Thales and Aristeia programmes co-financed by EU-ESF and the Greek NSRF, Greece; BSF-NSF and MINERVA, Israel; NCN and NAWA, Poland; La Caixa Banking Foundation, CERCA and AGAUR programs from Generalitat de Catalunya and PROMETEO and GenT Programmes Generalitat Valenciana, Spain; G\"{o}ran Gustafssons Stiftelse, Sweden; The Royal Society and Leverhulme Trust, United Kingdom; Eric and Wendy Schmidt Fund for Strategic Innovation, United States of America.

In addition, individual members wish to acknowledge support from Chile: Agencia Nacional de Investigaci\'on y Desarrollo (ANID FONDECYT reg. 1230987, FONDECYT 1230812, FONDECYT 1240864, Fondecyt 3240661, Fondecyt Regular 1240721); China: Chinese Ministry of Science and Technology (MOST-2023YFA1605700, MOST-2023YFA1609300), National Natural Science Foundation of China (NSFC - 12175119, NSFC 12275265); Czech Republic: Czech Science Foundation (GACR - 24-11373S), Ministry of Education Youth and Sports (ERC-CZ-LL2327, FORTE CZ.02.01.01/00/22\_008/0004632), PRIMUS Research Programme (PRIMUS/21/SCI/017); EU: H2020 European Research Council (ERC - 101002463); European Union: European Research Council (BARD No. 101116429, ERC - 948254, ERC 101089007), European Regional Development Fund (HE COFUND GA No.101081355, ERDF), European Union, Future Artificial Intelligence Research (FAIR-NextGenerationEU PE00000013), Marie Sklodowska-Curie Actions (GAP-101168829); France: Agence Nationale de la Recherche (ANR-21-CE31-0013, ANR-21-CE31-0022, ANR-22-EDIR-0002, ANR-24-CE31-0504-01); Germany: Deutsche Forschungsgemeinschaft (DFG - 469666862); China: Research Grants Council (GRF); Italy: Ministero dell'Università e della Ricerca (NextGenEU 153D23001490006 M4C2.1.1, NextGenEU I53D23000820006 M4C2.1.1, NextGenEU I53D23001490006 M4C2.1.1, SOE2024\_0000023); Japan: Japan Society for the Promotion of Science (JSPS KAKENHI  JP25H0063, JSPS KAKENHI JP22H01227, JSPS KAKENHI JP22H04944, JSPS KAKENHI JP22KK0227, JSPS KAKENHI JP24K23939, JSPS KAKENHI JP24KK0251, JSPS KAKENHI JP25H00650, JSPS KAKENHI JP25H01291, JSPS KAKENHI JP25K01011, JSPS KAKENHI JP25K01023); Norway: Research Council of Norway (RCN-314472); Poland: Ministry of Science and Higher Education (IDUB AGH, POB8, D4 no 9722), Polish National Science Centre (NCN 2021/42/E/ST2/00350, NCN OPUS 2023/51/B/ST2/02507, NCN OPUS nr 2022/47/B/ST2/03059, NCN UMO-2019/34/E/ST2/00393, UMO-2022/47/O/ST2/00148, UMO-2023/49/B/ST2/04085, UMO-2023/51/B/ST2/00920, UMO-2024/53/N/ST2/00869); Portugal: Foundation for Science and Technology (FCT); Spain: Agència de Gestió d'Ajuts Universitaris i de Recerca. (AGAUR - 2023 BP 00141), Generalitat Valenciana (ASFAE/2022/008), Ministry of Science and Innovation (RYC2019-028510-I, RYC2020-030254-I, RYC2021-031273-I, RYC2022-038164-I), Ministerio de Ciencia, Innovación y Universidades/Agencia Estatal de Investigaci\'on (EU NextGenerationEU (PRTR-C17.I1), PID2022-142604OB-C22); Sweden: Carl Trygger Foundation (Carl Trygger Foundation CTS 22:2312), Swedish Research Council (Swedish Research Council 2023-04654, VR 2021-03651, VR 2022-03845, VR 2022-04683, VR 2023-03403, VR 2024-05451), Knut and Alice Wallenberg Foundation (KAW 2018.0458, KAW 2022.0358, KAW 2023.0366); Switzerland: Swiss National Science Foundation (SNSF - PCEFP2\_194658); United Kingdom: The Binks Trust, Royal Society (NIF-R1-231091); United States of America: U.S. Department of Energy (ECA DE-AC02-76SF00515), John Templeton Foundation (John Templeton Foundation 63206), Neubauer Family Foundation.

%
%

% End of text imported from the .//acknowledgements/Acknowledgements.tex input file

%

%
%
%
\printbibliography
\clearpage
% ATLAS Collaboration author list
% Reference date of HMBS-2024-28 is 2025-10-24
% Author list last updated on date 06-APR-26
% Data extracted on 06-Apr-2026 for paper reference HMBS-2024-28
% at 5:42pm
 
\begin{flushleft}
\hypersetup{urlcolor=black}
{\Large The ATLAS Collaboration}

\bigskip

\AtlasOrcid[0000-0002-6665-4934]{G.~Aad}$^\textrm{\scriptsize 103}$,
\AtlasOrcid[0000-0001-7616-1554]{E.~Aakvaag}$^\textrm{\scriptsize 17}$,
\AtlasOrcid[0000-0002-5888-2734]{B.~Abbott}$^\textrm{\scriptsize 122}$,
\AtlasOrcid[0000-0002-0287-5869]{S.~Abdelhameed}$^\textrm{\scriptsize 84b}$,
\AtlasOrcid[0000-0002-1002-1652]{K.~Abeling}$^\textrm{\scriptsize 55}$,
\AtlasOrcid[0000-0001-5763-2760]{N.J.~Abicht}$^\textrm{\scriptsize 49}$,
\AtlasOrcid[0000-0002-8496-9294]{S.H.~Abidi}$^\textrm{\scriptsize 30}$,
\AtlasOrcid[0009-0003-6578-220X]{M.~Aboelela}$^\textrm{\scriptsize 45}$,
\AtlasOrcid[0000-0002-9987-2292]{A.~Aboulhorma}$^\textrm{\scriptsize 36e}$,
\AtlasOrcid[0000-0001-5329-6640]{H.~Abramowicz}$^\textrm{\scriptsize 155}$
% ...
\bigskip
\\

% ...
$^{17}$Department for Physics and Technology, University of Bergen, Bergen; Norway.\\
% ...
$^{30}$Physics Department, Brookhaven National Laboratory, Upton NY; United States of America.\\
% ...
$^{36}$$^{(a)}$Facult\'e des Sciences Ain Chock, Universit\'e Hassan II de Casablanca;$^{(b)}$Facult\'{e} des Sciences, Universit\'{e} Ibn-Tofail, K\'{e}nitra;$^{(c)}$Facult\'e des Sciences Semlalia, Universit\'e Cadi Ayyad, LPHEA-Marrakech;$^{(d)}$LPMR, Facult\'e des Sciences, Universit\'e Mohamed Premier, Oujda;$^{(e)}$Facult\'e des sciences, Universit\'e Mohammed V, Rabat;$^{(f)}$Institute of Applied Physics, Mohammed VI Polytechnic University, Ben Guerir; Morocco.\\
% ...
$^{45}$Physics Department, Southern Methodist University, Dallas TX; United States of America.\\
% ...
$^{49}$Fakult\"{a}t Physik, Technische Universit{\"a}t Dortmund, Dortmund; Germany.\\
% ...
$^{55}$II. Physikalisches Institut, Georg-August-Universit\"{a}t G\"ottingen, G\"ottingen; Germany.\\
% ...
$^{84}$$^{(a)}$Khalifa University of Science and Technology, Abu Dhabi;$^{(b)}$New York University Abu Dhabi, Abu Dhabi;$^{(c)}$United Arab Emirates University, Al Ain;$^{(d)}$University of Sharjah, Sharjah; United Arab Emirates.\\
% ...
$^{103}$CPPM, Aix-Marseille Universit\'e, CNRS/IN2P3, Marseille; France.\\
% ...
$^{122}$Homer L. Dodge Department of Physics and Astronomy, University of Oklahoma, Norman OK; United States of America.\\
% ...
$^{155}$Raymond and Beverly Sackler School of Physics and Astronomy, Tel Aviv University, Tel Aviv; Israel.\\
% ...
$^{a}$ Also at Affiliated with an institute formerly covered by a cooperation agreement with CERN.\\
$^{b}$ Also at An-Najah National University, Nablus; Palestine.\\
$^{c}$ Also at Borough of Manhattan Community College, City University of New York, New York NY; United States of America.\\
$^{d}$ Also at Center for Interdisciplinary Research and Innovation (CIRI-AUTH), Thessaloniki; Greece.\\
$^{e}$ Also at Centre of Physics of the Universities of Minho and Porto (CF-UM-UP); Portugal.\\
$^{f}$ Also at CERN, Geneva; Switzerland.\\
% ...

\end{flushleft}

% Created with Glance <Atlas.Glance@cern.ch>

%

%
%

%

%
%
%
%

\end{document}